\documentclass[twocolumn, dvipsnames]{aastex7}

\usepackage{xspace}
\usepackage{color}
\usepackage{comment}

\newcommand{\Mr}{$M_r$\xspace}

\newcommand{\pr}{$r$\xspace}
\newcommand{\sig}{$\sigma_{12}$\xspace}

\shorttitle{Variable AGN in Voids}
\shortauthors{Aradhey et al.}

\begin{document}

\title{Quantifying the Active Galactic Nuclei Fraction in Cosmic Voids via Mid-Infrared Variability}

\author[orcid=0009-0003-7308-9413]{Anish S. Aradhey}
\affiliation{Department of Physics and Astronomy, University of North Carolina at Chapel Hill, 120 E. Cameron Ave., Phillips Hall CB3255, Chapel Hill, NC 27599, USA}
\email[show]{aaradhey@unc.edu}

\author[orcid=0000-0002-2441-1619]{Anca Constantin}
\affiliation{Department of Physics and Astronomy, James Madison University, MSC 4502, 901 Carrier Drive, Harrisonburg, Virginia 22807, USA}
\email{constaax@jmu.edu}

\author[orcid=0000-0001-7416-9800]{Michael S. Vogeley}
\affiliation{Department of Physics, Drexel University, 32 S. 32nd Street, Philadelphia, PA 19104, USA}
\email{vogeley@drexel.edu}

\author[orcid=0000-0002-9540-546X]{Kelly A. Douglass}
\affiliation{Department of Physics and Astronomy, University of Rochester, 206 Bausch and Lomb Hall, P.O. Box 270171, Rochester, NY 14627, USA}
\email{kellyadouglass@rochester.edu}

\begin{abstract}

Observations and theoretical simulations suggest that the large scale environment plays a significant role in how galaxies form and evolve and, in particular, whether and when galaxies host an actively accreting supermassive black hole in their center (i.e., an Active Galactic Nucleus, or AGN). One signature of AGN activity is luminosity variability, which appears in the mid-infrared (mid-IR) when circumnuclear dust reprocesses UV and optical photons from the AGN accretion disk. We present here a suite of constraints on the fraction of AGN activity in the most underdense regions of the universe (cosmic voids) relative to the rest of the universe (cosmic walls) by using $\sim$12 years of combined multi-epoch data from AllWISE and NEOWISE to quantify mid-IR variability. We find clear evidence for a larger mid-IR variability-AGN fraction among high and moderate-luminosity void galaxies compared to their wall counterparts.  We also show that mid-IR variability identifies a rather large and unique population of AGNs, the majority of which have eluded detection using more traditional AGN-selection methods such as single-epoch mid-IR color selection. The fraction of these newly-recovered AGNs is larger among galaxies in voids, suggesting once again more prolific AGN activity in the most underdense large scale structures of the universe.  

\end{abstract} 
 
 
\section{Introduction}

\subsection{Galaxies in Void Environments}

Galaxy redshift surveys have revealed that galaxies display a web-like distribution of dense clusters connected by filaments that surround vast underdense regions known as cosmic voids \citep[see, e.g.][]{giovanelli_redshift_1991, bond_web_1996}. The void regions host galaxies (hereafter void galaxies) that appear distinct from their counterparts in more crowded regions (hereafter wall galaxies) in many ways: void galaxies are fainter and bluer, have surface brightness profiles more similar to those of late-type systems, higher specific star formation rates, and their mass and luminosity functions clearly are shifted towards lower characteristic mass and fainter magnitudes \citep{ rojas_photometric_2004, Rojas05, hoyle_luminosity_2005, ceccarelli_large-scale_2008, ceccarelli_impact_2022}.

These differences, along with studies of their central emission activity, suggest that, compared to their wall counterparts, void galaxies follow a different evolutionary path \citep[e.g.,][]{constantin_active_2008, ceccarelli_impact_2022, Dominguez-Gomez2023}. 
The evolution of these extremely isolated galaxies within cosmic voids is indeed expected to be different from those of their wall counterparts because of supposedly fewer processes such as stripping, harassment, and galaxy mergers, which are believed to preferentially influence galactic evolution and activity in crowded environments \citep[e.g.,][]{lin_dry_2010, lin_starrs_2014, jian_group_2017, pearson_merger_2024}.  The assumed lack of interactions within cosmic voids implies, to a first approximation, that the void galaxies' traits should resemble, or be more dominated by, those associated with a secular evolution, driven by internal slow, long-term, and gradual processes, rather than those rapidly acquired via external interactions. 
Thus, studying void galaxy properties is important for further constraining the galactic ``nature versus nurture" investigations, which have produced rather uncertain conclusions to date \citep[e.g.,][]{gunn_infall_1972, larson_evolution_1980, di_matteo_energy_2005, vanderwel_dependence_2008, peng_mass_2010, schawinski_zoo_2014, davies_host_2017, momose_Lya_2021, fahey_structural_2025}.

\subsection{Active Galactic Nuclei in Voids}

Within the debate over the extent to which the environment of a galaxy determines its evolution, a particular topic of contention is the connection between galaxy mergers and the accretion of matter onto the supermassive black holes at galaxy centers, i.e., active galactic nuclei (AGNs). This connection is both supported \citep[e.g.,][]{keel_interaction_1985, alonso_interaction_2007, ellison_interactions_2011, satyapal_mergers_2014, ellison_definitive_2019, gao_mergers_2020, mishra_lower_2020} and rejected \citep[e.g.,][]{grogin_bulge_2005, hopkins_we_2014, mechtley_mergers_2016, garland_most_2023} by an extensive body of observational and theoretical studies. A strong merger-AGN connection implies that AGNs should be most common among galaxies that experience frequent mergers, thus predicting a higher AGN fraction in wall galaxies compared to void galaxies. Studying and comparing the AGN fraction in void and wall galaxies may therefore be one of the strongest tests of the merger-AGN connection.

Previous studies investigating AGN abundance in cosmic voids reveal somewhat mixed results that depend on host galaxy properties. \citet{constantin_active_2008} show that AGNs are more common in voids than walls for moderately luminous and massive galaxies ($M_r \sim -20$, $\log M/M_\odot < 10.5$), while the AGN fraction remains comparable for the brighter hosts ($M_r < -20$) of the different environments. \citet{argudo_dependence_2018} find a higher AGN fraction in denser large scale regions compared to void regions for quenched and red isolated galaxies. \citet{ceccarelli_impact_2022} confirm that AGNs are significantly more common in voids than in walls for luminous galaxies ($-23 < M_r < -20.5$) regardless of whether AGN activity is identified via optical emission line diagnostics or mid-infrared (mid-IR) color selection.
In contrast, both \citet{liu_spectral_2015} (comparing void and wall environments) and \citet{amiri_role_2019} (comparing voids and rich galaxy clusters) find that the AGN fraction is weakly, if at all, affected by the local galaxy density.

This apparent lack of consensus and the fact that the majority of these studies
rely exclusively on optical emission line AGN selection techniques argue for a clear need for further study of AGN phenomenology in void galaxies using additional AGN identification methods.

\subsection{Mid-infrared Variability}

The identification of signs of nuclear BH accretion in void galaxies is particularly challenging, due to their tendencies to be fainter and host higher star formation rates, which can skew standard AGN identification methods such as optical emission line diagnostics \citep[e.g.,][]{morse_gaseous_1996, veilleux_spectroscopic_2002} and mid-IR color selection \citep{stern_selection_2005}. Void AGN, with typical luminosities that are lower than those associated with their host star formation, are therefore at higher risk of being overlooked.

A novel and promising method for identifying AGN in void galaxies is by quantifying the degree of mid-IR luminosity variability. AGN are known to show variable emission at all wavelengths (e.g., \citealp{ulrich_variability_1997, sartori_model_2018}), possibly due to instabilities in the accretion disk \citep[e.g.,][]{ruan_evidence_2014}. If the accretion is obscured by dust, the variable UV photons may then be reprocessed and produce variability in the mid-IR. Past mid-IR variability studies of AGN using observations from \textit{Spitzer} \citep{kozlowski_mid-infrared_2010, kozlowski_quasar_2016, polimera_morphologies_2018} and \textit{WISE} \citep{secrest_low_2020, son_mid-infrared_2022} reveal a unique population of optically obscured, Compton-thick AGN, some of which are completely missed by optical emission line and/or mid-IR color diagnostics.

In this paper, we employ for the first time mid-IR variability to study AGN activity in void galaxies. We compare the fraction of void galaxies that show mid-IR variability as an indicator of AGN activity to the same fraction among galaxies in more dense environments, with the goal of testing the connection between large-scale environment and nuclear galactic activity.  
In Section~\ref{sec:data}, we describe our sample selection along with the multi-epoch mid-IR data. We present in Section~\ref{sec:method} our calculations of a suite of statistical variability metrics, which we combine to define four classes of variable galaxies in Section~\ref{sec:vardefs}. We discuss our results in Section~\ref{sec:results} and list our main conclusions in Section~\ref{sec:conclusion}.

\section{The Data}

\label{sec:data}

\subsection{The Void and Wall Galaxy Samples}

We employ in this study the sample of void and wall galaxies from \citet{douglass_updated_2023}, who provide an updated catalog of cosmic voids in the Sloan Digital Sky Survey Data Release 7 \citep[SDSS DR7;][]{SDSS7}, based on the NASA-Sloan Atlas (NSA) version {\tt\string v1\_0\_1} \citep{blanton_improved_2011} using a new implementation of the VoidFinder algorithm \citep{elad_voidfinder_1997, Hoyle02, VAST}.  This catalog provides improved handling of survey boundaries and therefore of the definition of void and wall galaxies.

To identify void regions within a galaxy redshift survey, VoidFinder first separates the galaxies into field and wall galaxies based on their local density: field galaxies are those with their third nearest neighbor further than $\overline{d}_3 + 1.5\sigma_{d_3}$, where $\overline{d}_3$ is the average distance to the third nearest neighbor for all galaxies within a volume-limited catalog, and $\sigma_{d_3}$ is the standard deviation of the distribution of third nearest neighbor distances in the sample; wall galaxies are those with their third nearest neighbor closer than this limit.  The wall galaxies are then placed on a three-dimensional grid with side lengths of 5~Mpc/$h$. From the center of each empty grid cell, VoidFinder grows a sphere, expanding its radius and shifting its center until it is bounded by four galaxies on its surface.  These spheres are then joined to form the voids, with each void defined by a maximal sphere (the largest sphere in that void) and a number of smaller spheres that each overlap the maximal sphere by at least 50\% of their volume.  The maximal spheres are required to be at least 10~Mpc/$h$ in radius, and they cannot overlap each other by more than 10\% of their volume.  For more details on VoidFinder, see \citet{douglass_updated_2023}.

This algorithm assigns each of the galaxies one of three large-scale environment classifications: ``void'' for the galaxies falling within the defined void regions, ``wall'' otherwise, or ``edge'' when the wall galaxies fall too close to the survey boundary, i.e., where VoidFinder's ability to identify voids degrades. 
We consider here only the galaxies with the ``void'' and ``wall'' labels in order to focus on 
galaxies with confident VoidFinder classifications. The \citet{douglass_updated_2023} void catalog classifies about 52\% of the 641,409 galaxies in the entire NSA Catalog as residing in either void or wall environments, including 86,891 sources in voids and 245,778 sources in walls.

To better define our samples we make some small cuts on both redshift ($z$) and NSA $r$-band absolute magnitude (\Mr). The \citet{douglass_updated_2023} void catalog extends to $z = 0.114$, so we enforce a slightly more conservative cut of $z \leqslant 0.11$. Additionally, we enforce absolute magnitude cuts of $-24 \leqslant M_r \leqslant -10$. 
These cuts discard $\sim$2\% of our sample (see Table \ref{tbl:cuts}); they are simply intended to clearly bound the range of absolute magnitudes in our sample.
Together, these conditions return 84,075 void galaxies and 241,855 wall galaxies, or 97\% of the void galaxies and 98\% of the wall galaxies initially selected from the NSA Catalog, respectively.

\begin{figure}
    \centering
    \includegraphics[width=1.0\linewidth]{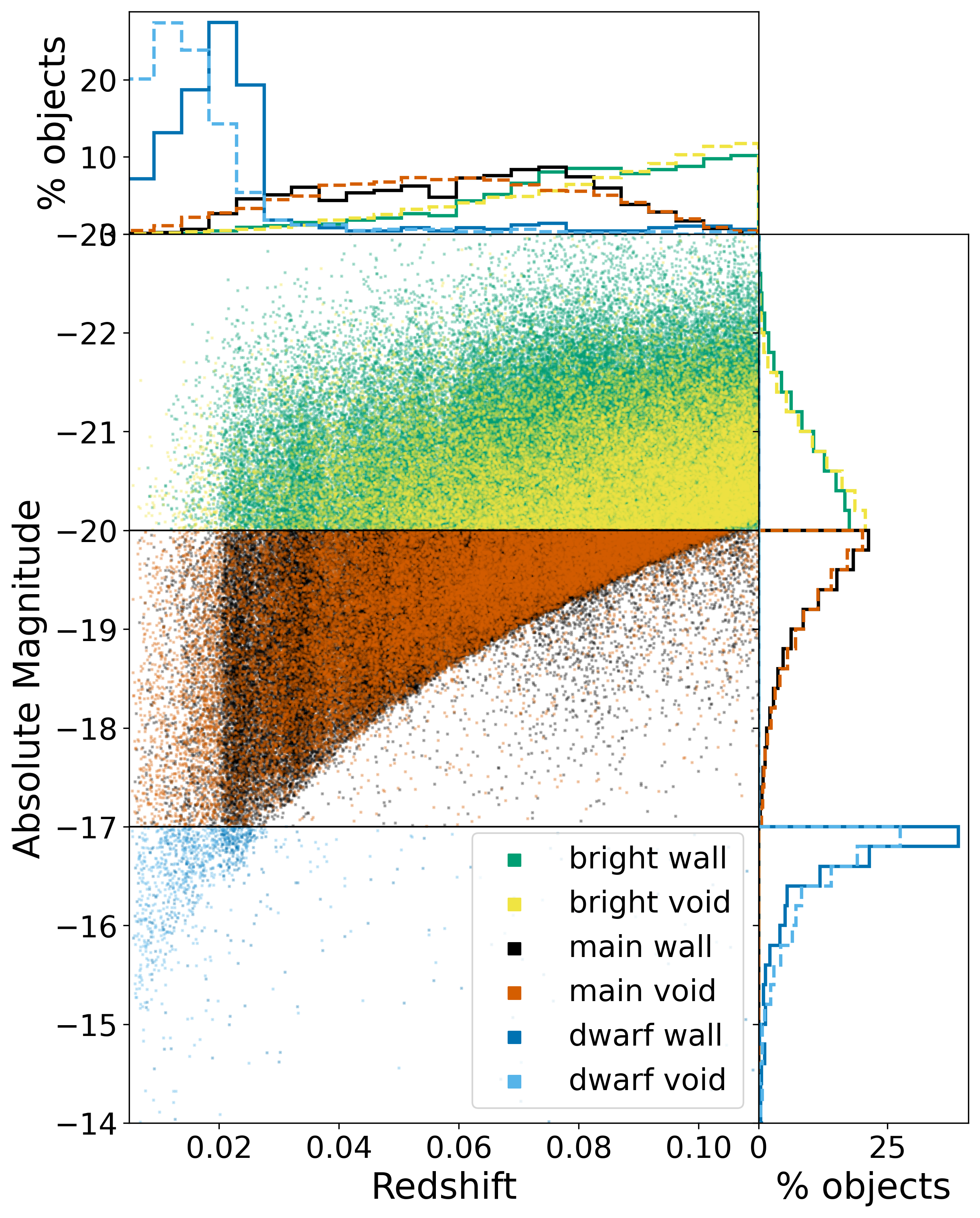}
    \caption{Void and wall subsample definitions based on their redshift $z$ -- $r$-band  magnitude $M_r$ distributions. We only consider sources with an absolute magnitude between $-24$ and $-10$ and $z \leqslant 0.11$. ``Bright'' sources have $M_r \leqslant -20$, ``main'' sources have $-20 < M_r \leqslant -17$, and ``dwarf'' sources have $M_r > -17$, shown as horizontal black lines.}
    \label{fig:Mr_vs_Z}
\end{figure}

Figure~\ref{fig:Mr_vs_Z} presents the distributions of both void and wall galaxies in redshift $z$ and the $r$-band absolute magnitude \Mr. In order to control for potential luminosity biases, we define and separately work with three luminosity subsamples: ``dwarf'' galaxies ($M_r >-17$), ``main'' galaxies ($-17 \geqslant M_r > -20$), and ``bright'' galaxies ($M_r \leqslant -20$).

\subsection{Single- and Multi-epoch Mid-Infrared Data} \label{subsec:mid-IR_data}

To characterize the mid-IR variability of these galaxies, we match them to the public all-sky AllWISE Source Catalog\footnote{https://doi.org/10.26131/IRSA1} at NASA/IPAC Infrared Science Archive\footnote{http://wise2.ipac.caltech.edu/docs/release/allsky/} via the IRSA\footnote{ http://irsa.ipac.caltech.edu/Missions/wise.html} catalog tool. Within IRSA, we query for the closest WISE detection within 6 arcseconds of each void and wall galaxy position, and we require that the WISE matches are detected with signal-to-noise ratios SNR $\geqslant 5$ in both the 3.4$\mu$m and 4.6$\mu$m bands (W1 and W2, respectively), which are used to identify and classify these galaxies as AGN based on the W1$-$W2 color \citep[e.g.,][]{stern_mid-infrared_2012}, as well as to quantify possible time variability. 
To avoid selecting sources flagged as spurious detections of image artifacts, we follow the suggestion in the online AllWISE Source Catalog documentation\footnote{https://wise2.ipac.caltech.edu/docs/release/allwise/expsup/ \newline sec2\_2.html} and require all galaxies to have {\tt cc\_flags} $= 0$. We also require all galaxies to have a measurement in the W3 band in order to calculate their W2--W3 colors, although we do not enforce a SNR limit in this band because we do not use W3 to calculate our variability metrics.
A total of 78,478 void galaxies and 228,871 wall galaxies (93\% and 95\% of the $z$ and $M_r$-limited void and wall samples, respectively) satisfy these conditions and are therefore AllWISE matches.

\begin{figure}
    \centering
    \includegraphics[width=1.0\linewidth]{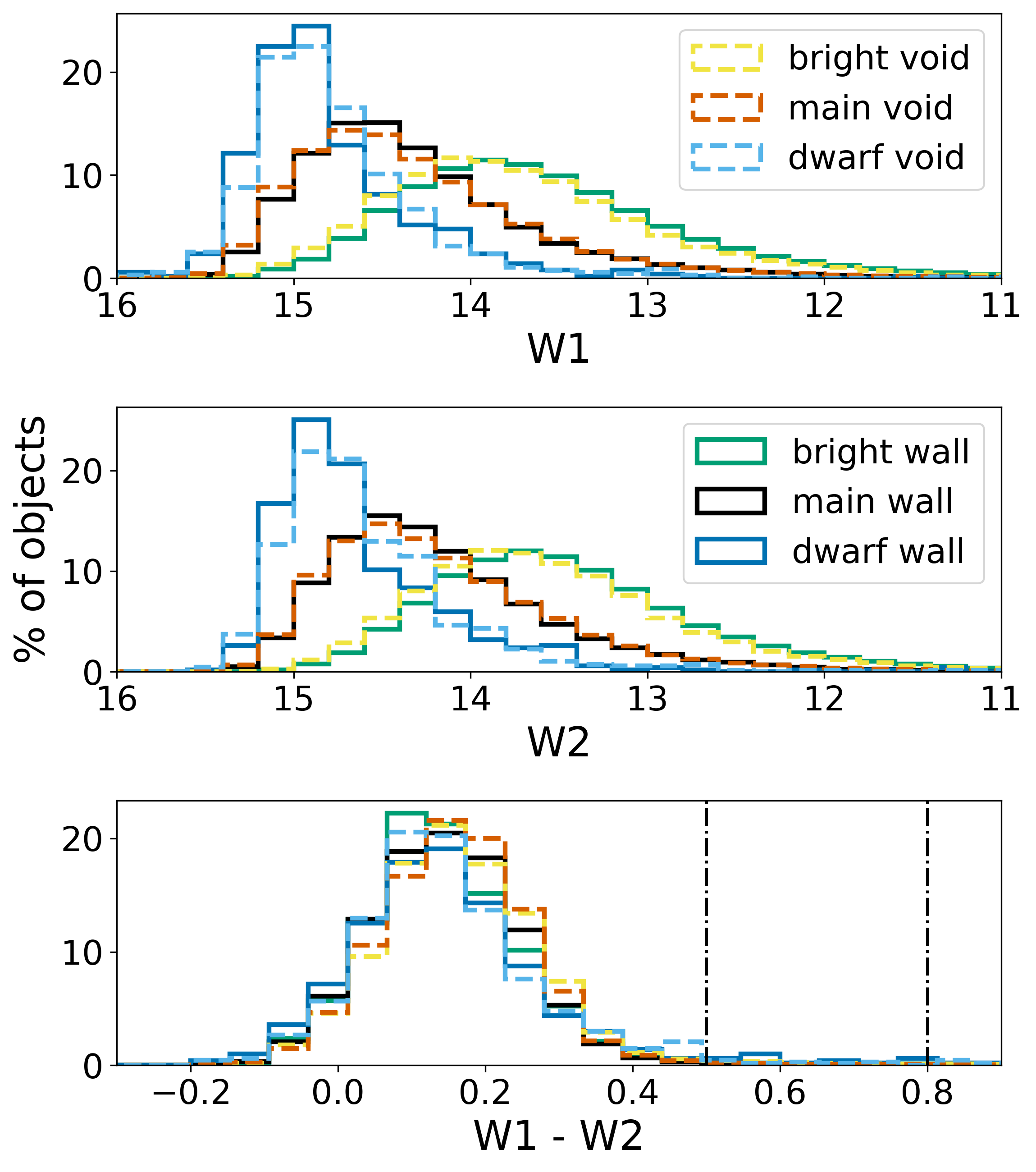}
    \caption{Histograms of W1 and W2 WISE magnitudes, along with the W1$-$W2 colors, shown separately for void galaxies (dashed lines) and wall galaxies (solid lines) in the three luminosity bins: Bright (yellow \& green), Main (red \& black), and Dwarf (light and dark blue).
    The vertical black dot-dashed lines illustrate the mid-IR AGN color cuts of W1$-$W2 $= 0.5$ \citep{ashby_spitzer_2009}, and W1$-$W2 $= 0.8$ \citep{stern_mid-infrared_2012}. Note the relative consistency between the distributions of W1 and W2 of the void and wall galaxies in each subsample. }
    \label{fig:WISE_hist}
\end{figure}

\begin{deluxetable}{lcc} 
\tablewidth{0 pt} 
\tablecaption{Mid-IR AGN fractions for Void and Wall Galaxies}
\tablehead{
\colhead{Sample} & \colhead{W1$-$W2$\geqslant0.5$} & \colhead{W1$-$W2$\geqslant0.8$} \\
\colhead{} & \colhead{\% (count)} & \colhead{\% (count)}
}
\startdata 
 Void & 1.39 ± 0.05 (941) & 0.44 ± 0.03 (297) \\ 
 Wall & 1.19 ± 0.02 (2490) & 0.36 ± 0.01 (763) \\ 
\\ 
Bright void & 1.72 ± 0.07 (559) & 0.44 ± 0.04 (143) \\ 
Bright wall & 1.38 ± 0.03 (1708) & 0.39 ± 0.02 (483) \\ 
\\ 
Main void & 1.02 ± 0.05 (353) & 0.39 ± 0.03 (134) \\ 
Main wall & 0.88 ± 0.03 (752) & 0.31 ± 0.02 (265) \\ 
\\ 
Dwarf void & 4.3 ± 0.8 (29) & 3.0 ± 0.7 (20) \\ 
Dwarf wall & 6 ± 1 (30) & 3.0 ± 0.8 (15)
\enddata
\label{tbl:colors}
\tablecomments{Fractions of void and wall galaxies that show AGN-like mid-IR colors (i.e., W1$-$W2 $\geqslant 0.8$, and W1$-$W2 $\geqslant 0.5$ for a more lenient definition) as reported in the AllWISE Source Catalog.
The listed errors are $\pm 1 \sigma$ Poisson uncertainties. The void samples show significantly ($\geqslant 2.3 \sigma$) higher mid-IR AGN fractions than the wall samples, with the exception of the Bright galaxies for W1$-$W2 $\geqslant 0.8$, and of the Dwarf galaxies, where the results remain inconclusive, perhaps due to the large uncertainties in such small subsamples.}
\end{deluxetable}

Figure~\ref{fig:WISE_hist} presents the distributions of W1, W2, and W1$-$W2 color of our subsamples, and we quantify the differences in the distributions of the W1$-$W2 colors in Table~\ref{tbl:colors}, which lists the fraction of objects with single-epoch WISE color cuts of W1$-$W2 $\geqslant 0.5$ and W1$-$W2 $\geqslant 0.8$, corresponding to traditional \citep{ashby_spitzer_2009} and more conservative \citep{stern_mid-infrared_2012} definitions for identifying mid-IR AGN.

Consistent with previous findings by \citet{ceccarelli_impact_2022}, this comparison of the W1$-$W2 colors for the full void and wall galaxy samples reveals that mid-IR AGN behavior is significantly more prevalent within the underdense environments.  
To quantify the statistical significance of this and other trends presented in this study, we divide the difference between the unrounded void and wall AGN fractions by the larger of the two associated unrounded $1\sigma$ Poisson uncertainties.  
This result is significant at the $4.4 \sigma$ level using the W1$-$W2 $\geqslant 0.5$ color cut and at the $2.9 \sigma$ level for the W1$-$W2 $\geqslant 0.8$ cut. However, the trend in the AGN fractions varies between the luminosity sub-samples and the AGN definitions: when AGNs are defined by W1$-$W2 $\geqslant 0.5$, there is a significantly higher AGN fraction in void galaxies within the Bright subsample ($4.6 \sigma$ significance) and Main subsample ($2.5 \sigma$ significance); for the more stringent definition, W1$-$W2 $\geqslant 0.8$, the enhancement of AGN activity within voids is only significant within the Main subsample, at the $2.3 \sigma$ level.

Within the Dwarf luminosity subgroup there is no statistically significant difference between the void and wall AGN fractions regardless of the mid-IR AGN definition, possibly because the uncertainties are so large in such small subsamples.

While mid-IR AGN classifications can be achieved with single-epoch color measurements, multi-epoch measurements are needed to identify AGN candidate based on variability, which is the focus of our study. To obtain multi-epoch mid-IR photometry for each galaxy, we crossmatch the WISE single epoch matches to the void/wall galaxy catalog coordinates to the ALLWISE Multiepoch Photometry (ALLWISE MEP)\footnote{https://doi.org/10.26131/IRSA134} and the NEOWISE-R Single Exposure (L1b) Source Tables\footnote{https://doi.org/10.26131/IRSA144}, using a $3\arcsec$ search radius, where we again select only high quality measurements by requiring SNR $\geqslant 5$ in both bands and enforcing the suggestions given in \citet{son_mid-infrared_2022} and the online NEOWISE Single-exposure Source Database documentation\footnote{https://wise2.ipac.caltech.edu/docs/release/neowise/expsup/ \newline sec2\_3.html}: {\tt cc\_flags} $= 0$, {\tt qual\_frame} $> 0$, {\tt qi\_fact} $>0$, {\tt saa\_sep} $> 0$, and {\tt moon\_masked} $= 0$. The ALLWISE MEP measurements do not report SNR, so we estimate SNR in both bands by dividing each measurement's single-exposure profile-fit magnitude ({\tt w1mpro\_ep}, {\tt w2mpro\_ep}) by the associated single-exposure profile-fit photometric measurement uncertainty ({\tt w1sigmpro\_ep}, {\tt w2sigmpro}).  We consider NEOWISE single-epoch measurements through the NEOWISE 2023 Data Release on 2023 March 22.

Following the methodology outlined in \citet{secrest_low_2020}, we bin these multi-epoch measurements for each galaxy into 10-day bins and remove measurements that deviate by more than 3 median absolute deviations from the median magnitude of each bin. The AllWISE MEP set includes less than three bins, while the NEOWISE data make up the majority of most light-curve measurements. To robustly detect variability, we require all galaxies to have at least 4 epochs of data in their light curve,
which is consistent with the methodology of past studies of mid-IR AGN variability \citep{kozlowski_quasar_2016}.
Our multi-epoch data cuts return initial samples of 71,533 void sources (91\% of the void AllWISE matches) and 217,204 wall sources (95\% of wall AllWISE matches).

\begin{figure}
    \centering
    \includegraphics[width=1.0\linewidth]{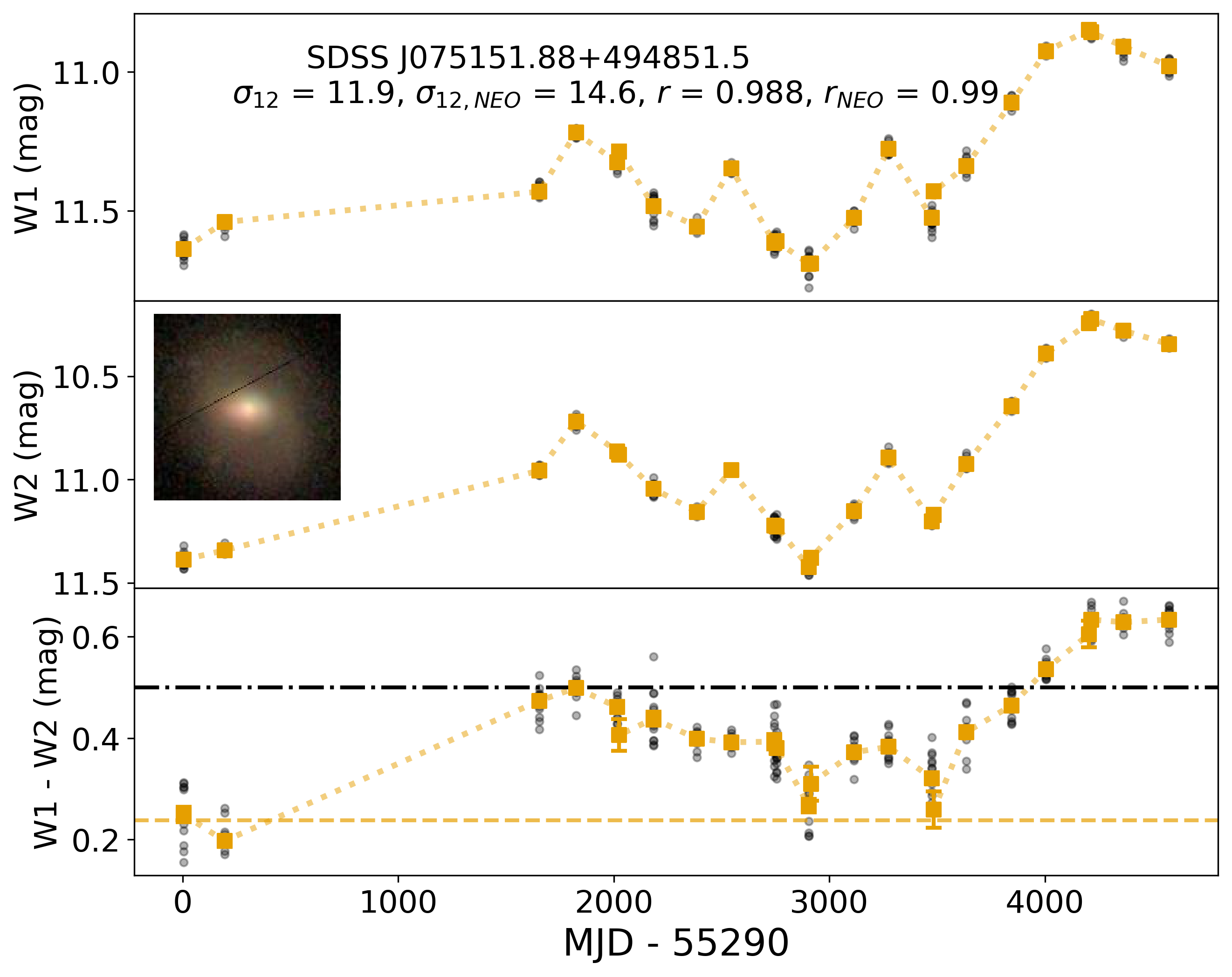}
    \caption{WISE light curves in W1, W2, and the W1$-$W2 color for one variable Bright galaxy identified in the void sample.
    Black dots are the individual measurements, orange squares show the arithmetic mean and the corresponding standard error \citep[$\sigma_e$,][]{son_mid-infrared_2022}. 
    The galaxy is labeled with its SDSS ID, Pearson \pr (see section \ref{pearsonr}), and \sig (see Section \ref{sigma12}) values. 
    In the bottom panel, the black horizontal dot-dashed line shows the W1$-$W2 $ = 0.5$ line for mid-IR AGN classification. The orange dashed line shows the galaxy's W1$-$W2 color as reported in the AllWISE Source Catalog. 
    The optical thumbnail image is from the SDSS DR7 Image List Tool.
    }
    \label{fig:LC_bright}
\end{figure}

Figure \ref{fig:LC_bright} illustrates an example of the WISE light curves in W1, W2, and the W1$-$W2 color for one variable Bright galaxy identified in the void sample, illustrating both the individual measurements and the median values in each bin, for all AllWISE MEP and NEOWISE observations.  Note how for this particular object the W1$-$W2 color changes significantly from non-AGN ($\sim 0.2$ in the MEP observations) to AGN-like ($> 0.5$) towards the end of the NEOWISE dataset. Also notable here is the redder-when-brighter behavior, which is consistent with the brightening being due to increasing AGN dominance in the host galaxy. This galaxy shows significant AGN-like variation despite not being classified as an AGN based on its single epoch AllWISE Source Catalog colors.

To quantify variability within just the NEOWISE portion of each galaxy's light curve, we also require that galaxies must have at least 4 epochs within this set alone; this additional condition allows us to calculate the Pearson \pr coefficient only for the NEOWISE measurements, $r_{\rm NEO}$, for about 95\% and 96\% of our initial void and wall galaxy samples, respectively. This yields our final sample of 276,959 sources: 67,758 void galaxies and 209,201 wall galaxies.

Table~\ref{tbl:cuts} lists the counts and percentages of void and wall galaxies after each cut, where the percentages are calculated out of the number of galaxies satisfying the previous cut. In general, the cuts select and discard very similar fractions of void and wall galaxies, especially within the Bright and Main luminosity subsamples. The fractions of void and wall galaxies selected by each cut deviate slightly within the Dwarf sample; this is expected because these sources tend to be fainter and therefore have noisier measurements.

\begin{deluxetable*}{lcccccc} 
\tablewidth{0 pt} 
\tablecaption{Effect of Each Filter on Sample Counts (\%)
}
\tablehead{
\colhead{Sample} & \colhead{NSA Catalog} & \colhead{Void or Wall} & \colhead{$M_{r}$ and $z$ } & \colhead{ALLWISE Source} & \colhead{$N_{\rm epochs} \geqslant 4$} & \colhead{$N_{\text{epochs, NEO}} \geqslant 4$} \\
\colhead{} & \colhead{} & \colhead{} & \colhead{limits} & \colhead{ cuts} & \colhead{} & \colhead{(final sample)}
}
\startdata 
All Galaxies & 641409 & 332669 (51.9) & 325930 (98.0) & 307349 (94.3) & 288737 (93.9) & 276959 (95.9) \\ 
\\ 
Void & -- & 86891 & 84075 (96.8) & 78478 (93.3) & 71533 (91.2) & 67758 (94.7) \\ 
Wall & -- & 245778 & 241855 (98.4) & 228871 (94.6) & 217204 (94.9) & 209201 (96.3) \\ 
\\ 
Bright void & -- & -- & 34467 & 32719 (94.9) & 32611 (99.7) & 32460 (99.5) \\ 
Bright wall & -- & -- & 130607 & 124341 (95.2) & 123988 (99.7) & 123561 (99.7) \\ 
\\ 
Main void & -- & -- & 46068 & 43264 (93.9) & 37860 (87.5) & 34626 (91.5) \\ 
Main wall & -- & -- & 107659 & 101991 (94.7) & 92253 (90.5) & 85137 (92.3) \\ 
\\ 
Dwarf void & -- & -- & 3540 & 2495 (70.5) & 1062 (42.6) & 672 (63.3) \\ 
Dwarf wall & -- & -- & 3589 & 2539 (70.7) & 963 (37.9) & 503 (52.2)
\enddata
\label{tbl:cuts}
\tablecomments{All percentages in a column are calculated out of the corresponding sample size in the previous column, i.e., the percentage out of the number of galaxies satisfying the previous cut. The rightmost column shows the final number of void and wall galaxies in the analysis.} 
\end{deluxetable*}

\section{Methodology: Statistical Variability Metrics} \label{sec:method}

We measure and compare the variability in the multi-epoch WISE data for each galaxy using three methods:
\begin{enumerate}

\item the Pearson correlation coefficient (\pr), which quantifies the coupling between the variation in the W1 and W2 bands \citep{kozlowski_quasar_2016, secrest_low_2020}, 

\item \sig, which measures the combined significance of variances in W1 and W2 by quantifying the degree to which the object deviates from the median variances in both bands \citep{kozlowski_quasar_2016}.

\item the percent of time the galaxy exhibits red AGN-like mid-IR colors ($p_{\rm AGN}$)

\end{enumerate} 

\subsection{Calculating Pearson r} \label{pearsonr}

\begin{figure}
    \centering
    \includegraphics[width=1.0\linewidth]{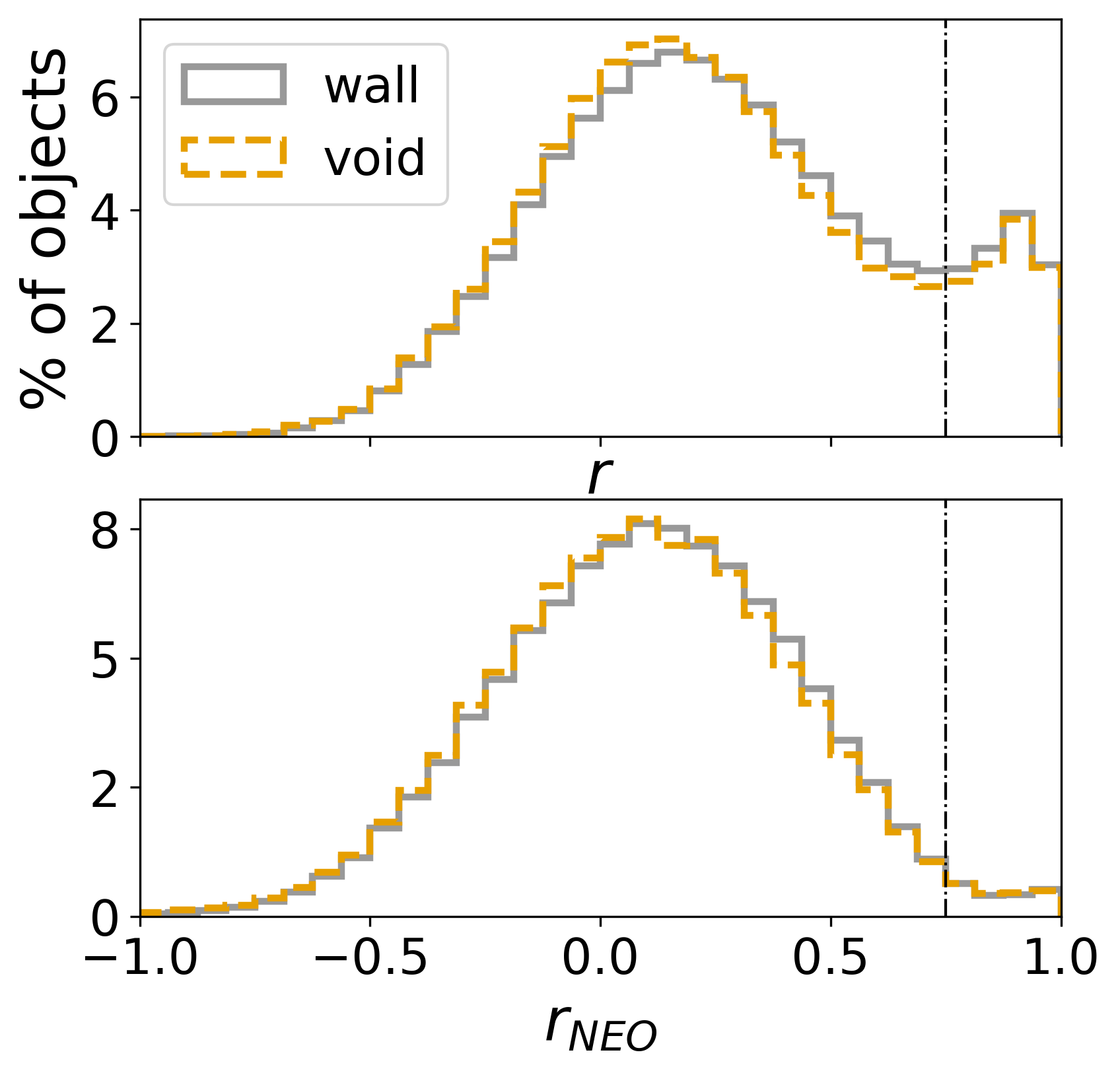}
    \caption{Distributions of the Pearson \pr correlation coefficients for void and wall sources calculated for the whole set of MEP and NEOWISE measurements (upper panel) and only for the NEOWISE data (lower panel). We choose a cut of Pearson $r > 0.75$ (dashed black vertical line) to select candidate variable sources.}
    \label{fig:Pr_whole}
\end{figure}

To quantify the coupling between variation in the W1 and W2 bands, we use Pearson's correlation coefficient \pr. With this measure, true variable sources should have W1 and W2 brightnesses following similar behaviors, translating into a high Pearson value \pr (i.e., $\approxeq$ 1; values of $-1$ and 0 would correspond to anticorrelated and uncorrelated, respectively). Variability criteria based on this coefficient included $r > 0.8$ (\citealt{kozlowski_quasar_2016}, \citealt{polimera_morphologies_2018}, both using Spitzer data) and $r > 0.4$ (\citealt{secrest_low_2020}, using WISE/NEOWISE data). 
We calculate \pr as:

\begin{equation}
r = \frac{C_{m_{1}, m_{2}}}{v_{m_{1}}v_{m_{2}}}
\label{eqn:r}
\end{equation}

where $C_{m_{1}, m_{2}}$ is the covariance between W1 and W2 for $N$ measurements of a single source:

\begin{equation}
C_{m_{1}, m_{2}} = \frac{1}{N - 1} \displaystyle\sum\limits_{i}^{N} (m_{1, i} - \langle m_{1} \rangle)(m_{2, i} - \langle m_{2} \rangle)
\label{eqn:Cm1m2}
\end{equation}

and $v_{m_{1}}$ and $v_{m_{2}}$ are the standard deviations as a function of the apparent magnitude in the band $X$ for that source:

\begin{equation}
v_{m_{X}}^{2} = \frac{1}{N - 1}\displaystyle\sum\limits_{i}^{N} (m_{X,i} - \langle m_{X} \rangle)^{2}
\label{eqn:vm}
\end{equation}

where $m_{X,i}$ is the mean magnitude of the source during epoch $i$, and $\langle m_X \rangle$ is the mean magnitude of the source in band $X$ across all epochs. 

\begin{figure*}
    \centering
    \includegraphics[width=1.0\linewidth]{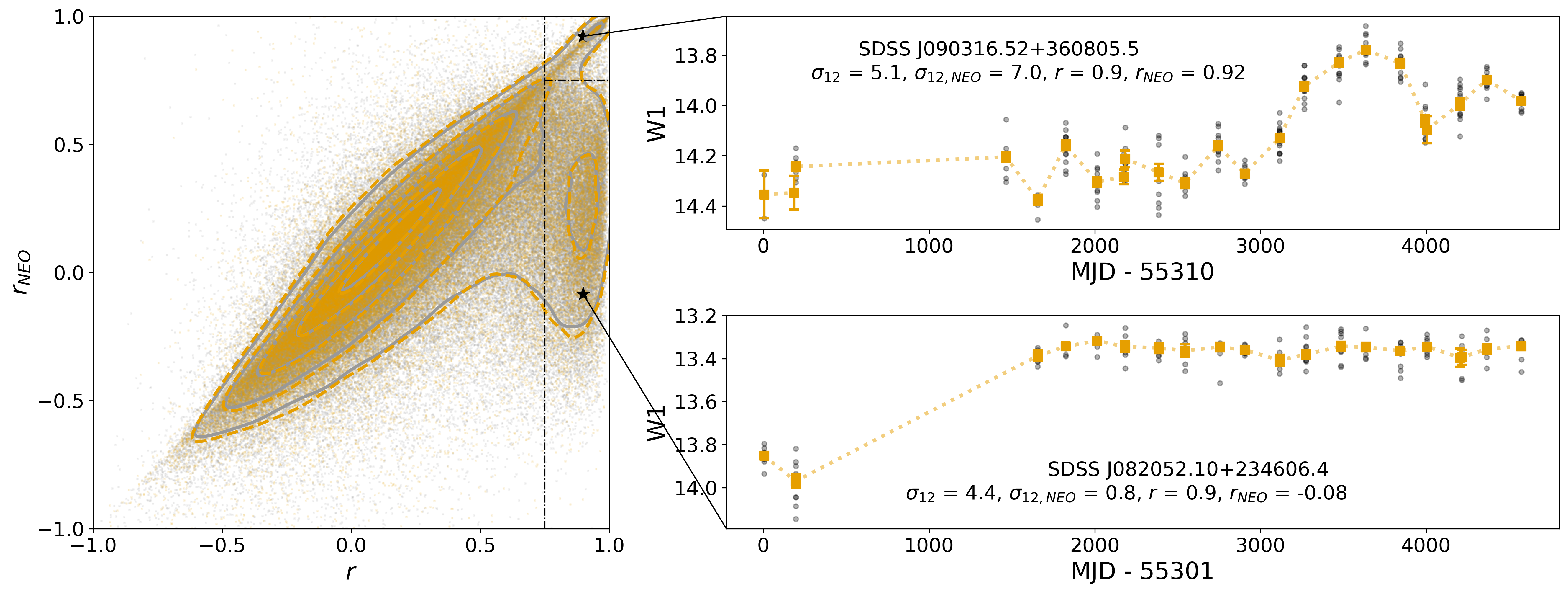}
    \caption{Distribution of the Pearson \pr coefficient calculated using all data ($x$-axis) and only NEOWISE data ($y$-axis) for the void (orange, dashed density contours) and wall (gray, solid density contours) galaxy samples. Colored dots show the location of individual galaxies, and the density contours correspond to factors of $n$ of the total number of objects in each class, where $n =$ 0.1, 0.2, 0.3, 0.5, 0.7, and 0.9 starting from innermost contour. The black stars show the locations of two void galaxies whose light curves are inset. The vertical dashed line at $r_{12} = 0.75$ and horizontal dashed line at $r_{12, NEO} = 0.75$ illustrate the cut criteria for variability.
    In the light curves, gray dots are the individual measurements, and orange squares show the arithmetic mean per 10-day epoch, with the corresponding standard error. The galaxies are labeled with their SDSS ID, Pearson \pr values (Section \ref{pearsonr}), and \sig values (Section \ref{sigma12}).
    }
    \label{fig:Pr_PrNEO}
\end{figure*}

We present in Figure~\ref{fig:Pr_whole} the distribution of \pr values for our entire void and wall galaxy samples, calculated for the whole set of MEP and NEOWISE measurements  (the \pr values) as well as only for the NEOWISE data (the $r_{\rm NEO}$ values). We expect the majority of sources to be non-variable galaxies, thus the distribution of \pr is dominated by contributions from noise, with the true variables being objects with a high positive correlation \citep[e.g.,][]{polimera_morphologies_2018, kozlowski_quasar_2016}. The distribution of Pearson \pr coefficients reveals a clear non-Gaussian tail at $r > 0.75$ while the $r_{\rm NEO}$ distribution appears more symmetric, with a less dramatic rise at $r_{\rm NEO} > 0.75$.  We quantify the differences between the Pearson \pr values between void and wall galaxies in Section~\ref{results}.

Upon visually inspecting the light curves of objects with very high Pearson \pr values, we find many that show variation only in the $\sim$3-year gap between the MEP and NEOWISE data, with no apparent variation within either set of measurements. These ``jumpy'' light curves also tend to show significant discrepancies between their \pr and $r_{\rm NEO}$ values, with $r > r_{\rm NEO}$. This discrepancy suggests that these light curves have misleadingly high \pr values because of the data gap between the MEP and NEOWISE observations rather than true variability.

We illustrate these discrepancies in Figure~\ref{fig:Pr_PrNEO}, with a direct comparison of the \pr and $r_{\rm NEO}$ coefficients, along with two examples of light curves for objects that show, respectively, consistency and inconsistency between the two Pearson \pr calculations. This comparison reveals that, while the majority of galaxies appear roughly along the  $r_{\rm NEO} = r$ line, there is an apparent cloud of ``jumpy'' sources with very high \pr values ($>0.75$) but low $r_{\rm NEO}$ values ($\leqslant 0.75$). This cloud contains 7,465 void sources (11.0\% of the void sample) and 24,357 wall sources (11.6\% of the wall sample). To avoid selecting such sources as candidate variables, we establish a ``simple'' criterion for variability involving both $r > 0.75$ and $r_{\rm NEO} > 0.75$, thus excluding sources with discrepancies between the two coefficients (see Section \ref{sec:vardefs}).

\subsection{Calculating \sig} \label{sigma12}

\begin{figure}
    \centering
    \includegraphics[width=1.0\linewidth]{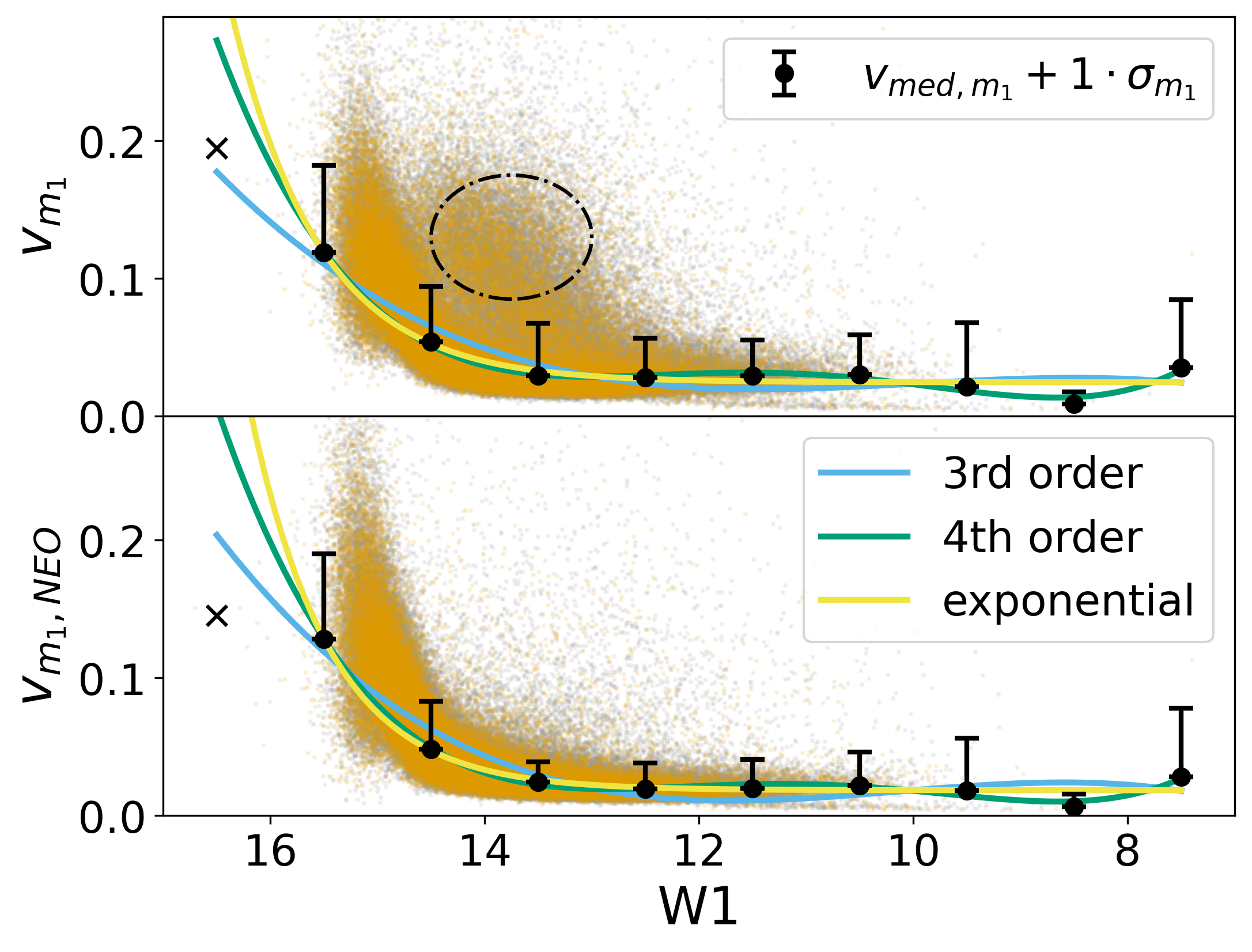}
    \caption{$v_{m_1}$ as a function of W1 when calculated using all light curve epochs (top panel) and just using NEOWISE epochs (bottom panel). Individual wall and void galaxies are shown as gray and orange dots, respectively. Black points show the median $v_{{\rm med}, {m_1}}$ and $v_{{\rm med}, {m_1}, {\rm NEO}}$ values in 1.0 mag-wide bins for all galaxies, placed at the center of their respective bins. Colored lines show three different functions (exponential, 3rd order polynomial, and 4th order polynomial) fit to $v_{{\rm med}, {m_1}}$. Black error bars show the $1\sigma_{m_1}$ dispersion above the median values for the respective apparent magnitude bins.
    Black $\times$ symbols show the ignored $v_{{\rm med}, {m_1}}$ value of the faintest bin. The black dash-dotted ellipse indicates a ``cloud'' of objects that have unusually high $v_{m_1}$ values for their apparent magnitude range, which disappears when considering only the $v_{m_1, {\rm NEO}}$ values. As expected, $v_{m_1}$ and $v_{{\rm med}, {m_1}}$ values increase systematically for the faintest galaxies. We find that the exponential (yellow) function best captures this trend. Galaxies with $v_{m_X}$ significantly above this trend line in both bands are assigned higher \sig values.}
    \label{fig:linefit}
\end{figure}

We follow the methodology of \citet{kozlowski_quasar_2016} and \citet{polimera_morphologies_2018} for identifying variable sources by supplementing the Pearson \pr criterion with a cut on \sig, which expresses the joint significance of the variability of the source in both W1 and W2 bands.
\sig has the potential to separate truly variable sources from the majority of sources in the field by identifying a new variability significance threshold; this threshold is based on the distribution of $v_{m_{X}}$ (i.e., the standard deviation of the W1 or W2 apparent magnitude for the whole light curve, as used in calculating the Pearson \pr coefficient; see Equation~\ref{eqn:vm}) which is dominated by contributions from noise rather than true variability. We define \sig for each galaxy as: 

\begin{equation}
\sigma_{12}^2 = \left( \frac{v_{m_1} - v_{{\rm med}, m_1}}{\sigma_{m_1}} \right)^2 + \left( \frac{v_{m_2} - v_{{\rm med}, m_2}}{\sigma_{m_2}} \right)^2
\label{eqn:sig12}
\end{equation}

Here, $v_{{\rm med}, m_{X}}$ is the median value of $v_{m_{X}}$ (see Equation~\ref{eqn:vm}, for the Pearson \pr coefficient) calculated for 1.0-mag-wide apparent magnitude bins 
in each of the W1 and W2 bands, regardless of large-scale environment. $\sigma_{m_{X}}$ is the standard deviation of $v_{m_{X}}$ values around the median $v_{{\rm med}, m_{X}}$ in each of these apparent magnitude bins.  
The dispersion of the $v_{m_{X}}$ values around the median ($\sigma_{m_{X}}$) in each apparent magnitude bin is calculated as:

\begin{equation}
\sigma_{m}^{2} = \frac{1}{N - 1}\displaystyle\sum\limits_{i}^{N} (v_{m_{Xi}} - v_{{\rm med},m})^{2}
\label{eqn:sig-m}
\end{equation}

where $N$ is the number of sources in each apparent magnitude bin. We experimented with several bin sizes (0.1 mag, 0.5 mag, and 1.0 mag) and chose the 1.0 mag bin width for simplicity and to minimize the number of bins with very low galaxy counts.

Equation~\ref{eqn:sig12} shows that \sig expresses the significance of a galaxy's variability in each band added in quadrature. Thus, the \sig values are always positive, with a higher \sig value indicating that a galaxy's variability is more significant compared to other galaxies in the sample with similar apparent magnitudes.
Past studies \citep{kozlowski_quasar_2016, polimera_morphologies_2018} have considered variable sources to be those with $\sigma_{12}> 2$, $>3$, or $>4$ , with higher thresholds being more conservative.

Before calculating \sig, we first examine the distribution of $v_{m_1}$ as a function of W1 for all void and wall galaxies for calculations of $v_{m_1}$ that include both MEP and NEOWISE light curve epochs (top panel of Figure~\ref{fig:linefit}) and only the NEOWISE epochs (bottom panel of Figure~\ref{fig:linefit}), along with the median $v_{{\rm med},m_{1}}$ and the $\sigma_{m_{1}}$ values calculated for each bin. As expected, $v_{m_1}$ increases dramatically at W1 $> 14$ because the distribution of $v_{m_1}$ values becomes statistically dominated by contributions from photometric noise \citep{kozlowski_quasar_2016}, which is more pronounced for fainter galaxies.
This trend is also apparent in the values of $v_{{\rm med},{m_1}}$ (black points) and $\sigma_{m_1}$ (black upper error bars).

In order to produce smooth distributions of \sig values,
rather than use the $v_{{\rm med}, X}$ and $\sigma_{m_{X}}$ values calculated in each individual bin,
we fit the $v_{{\rm med}, X}$ and $\sigma_{m_{X}}$ values from all bins as a function of their respective apparent magnitudes. We find that a three-parameter exponential fit, with the faintest bin ignored, provides the most adequate representation of the data with the least number of parameters. While this approach reduces confidence in our conclusions about the Dwarf subsample of galaxies, we find that among the Dwarf galaxies our results remain statistically insignificant regardless of the details of our \sig calculation, simply because of the very low sample size. 

\begin{figure*}
    \centering
    \includegraphics[width=1.0\linewidth]{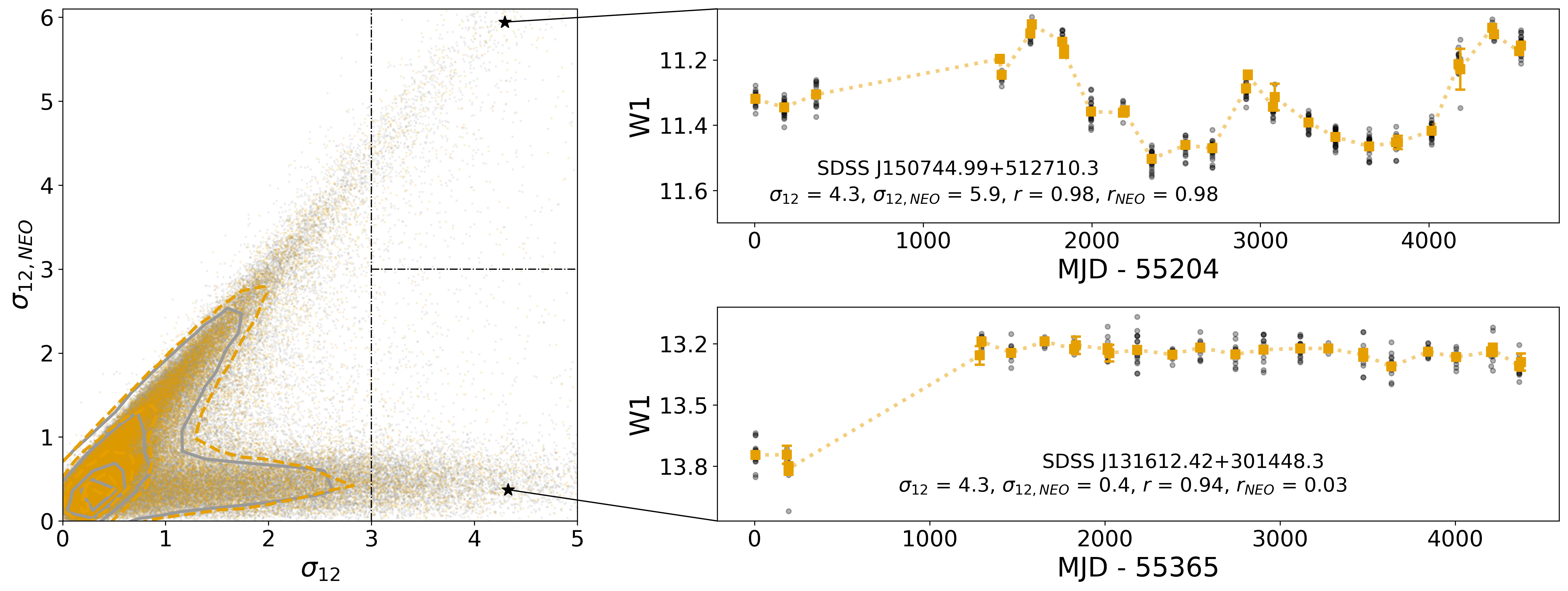}
    \caption{Distribution of the joint significance of variability (\sig) calculated using all data ($x$-axis) and only NEOWISE data ($y$-axis) for the void (orange) and wall (gray) galaxy samples. Colored dots show the location of individual galaxies, and the density contours correspond to factors of $n$ of the total number of objects in each class, where $n =$ 0.1, 0.2, 0.3, 0.5, 0.7, and 0.9 starting from innermost contour. The black stars show the location of two
    void galaxies whose light curves are inset. The black dashed lines illustrate cut criteria for variability corresponding to $\sigma_{12} > 3$ and $\sigma_{12, NEO} > 3$, where $\sigma_{12, NEO}$ is calculated using only NEOWISE light curve epochs.
    In the inset light curves, black dots are the individual measurements, and orange squares show the arithmetic mean per 10-day epoch and the corresponding standard error ($\sigma_e$) as defined by \citet{son_mid-infrared_2022}. The galaxies are labeled with their SDSS ID, Pearson \pr values (see Section \ref{pearsonr}), and \sig values (see Section \ref{sigma12}).
    }
    \label{fig:sig12_vs_NEO}
\end{figure*}

We explore in Figure~\ref{fig:sig12_vs_NEO} 
the effects of calculating \sig using all available epochs versus only the NEOWISE data, by comparing against each other both \sig and $\sigma_{12, {\rm NEO}}$ values, for void and wall galaxies. 
As expected, the vast majority of objects have $\sigma_{12} < 2$. While the \sig and $\sigma_{12, {\rm NEO}}$ metrics are positively correlated, there is a significant fraction of objects with a much higher \sig than $\sigma_{12, {\rm NEO}}$, creating a horizontal lower wing for $0 < \sigma_{12, {\rm NEO}} \leqslant 1$. We find 3,219 void sources (4.8\% of the void sample) and 10,164 wall sources (4.9\% of the wall sample) that are located in this wing (i.e. $\sigma_{12} > 2$ and $\sigma_{12, {\rm NEO}} \leqslant 1$). These lower wing objects are likely the ``jumpy'' ones, with no apparent variation within those two separate sets of measurements (see examples of light curves in insets), very similar to what we found when examining light curves that showed high Pearson \pr values. 
Since the NEOWISE epochs make up the majority of the galaxy light curves, the low $\sigma_{12, {\rm NEO}}$ values of these objects indicate they are probably not true variables.
Therefore, to classify a galaxy as variable, we must consider the \sig values calculated for both the whole multi-epoch data and only NEOWISE.
Section \ref{sec:vardefs} presents our criterion for variability based on \sig and $\sigma_{12, {\rm NEO}}$.

\subsection{$p_{\rm AGN}$: The Fraction of Observed Time with Mid-IR AGN Colors} \label{pAGN-def}

\begin{figure}
    \centering
    \includegraphics[width=1.0\linewidth]{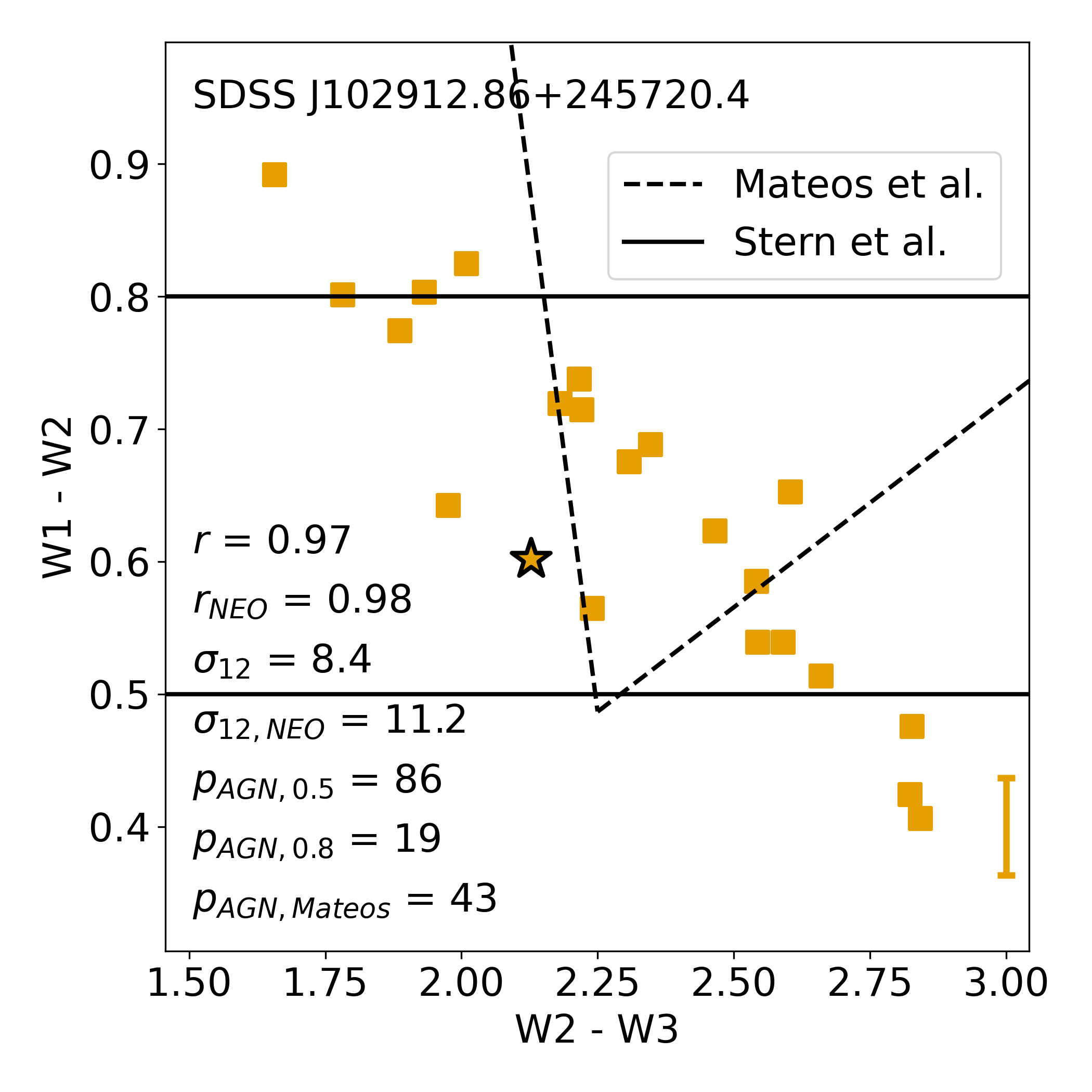}
    \caption{WISE colors of the averaged epochs in the light curve of one variable void galaxy in our sample, SDSS J102912.86+245720.4. The orange star shows the galaxy's W1$-$W2 and W2$-$W3 colors as reported in the AllWISE Source Catalog, while the orange squares show the colors in the averaged epochs of NEOWISE and MEP data. The error bar in the bottom right corner illustrates the average error for each epoch, as defined by \citet{son_mid-infrared_2022}.
    The black horizontal lines correspond to the AGN color cuts by \citet{stern_mid-infrared_2012}, and the black dashed line shows the AGN wedge cut from \citet{mateos_using_2012}. Note the significant variation from the AllWISE Source Catalog values.}
    \label{fig:GIF}
\end{figure}

Another way of quantifying differences in the mid-IR variability, and thus in the AGN activity between the void and wall environments, is by comparing the fraction of time that these galaxies appear AGN-like in their mid-IR colors during the WISE multi-epoch observations.  This measure should be more sensitive to AGN activity than simply using the colors reported in the AllWISE Source Catalog because it includes galaxies whose colors change over time and are expected to display AGN-like colors at epochs other than the one-time observation.  
We calculate here the fraction of each galaxy's light curve epochs that have W1$-$W2 values $\geqslant 0.5$ and $\geqslant 0.8$,
$p_{\text{AGN,0.5}}$ and $p_{\text{AGN,0.8}}$, along with the fraction of void and wall galaxies in each luminosity subsample that spent more than half of their time (as observed during the WISE/NEOWISE light curves) showing AGN-like mid-IR colors (i.e. $p_{\text{AGN,0.5}} > 50$ and $p_{\text{AGN,0.8}} > 50$).

Because its calculation involves taking a simple fraction, the $p_{\text{AGN}}$ method for AGN identification may seem at first more vulnerable than the previous two metrics to photometic artifacts, such as cosmic ray hits. However, we reduce the effect of such artifacts by calculating $p_{\text{AGN}}$ using the binned light curve epochs described in Section \ref{subsec:mid-IR_data}, each of which is the arithmetic mean of, on average, $\sim$9 individual measurements in the AllWISE MEP and NEOWISE data sets, with outliers discarded.

To demonstrate the usefulness of the $p_{AGN}$ parameter, we show in Figure~\ref{fig:LC_bright} one example of a binned light curve for the void galaxy SDSS J075151.88+494851.5.
This galaxy has W1$-$W2 $< 0.5$ in its single-epoch AllWISE measurements and would not be classified as a mid-IR AGN based this measurement.
However, Figure~\ref{fig:LC_bright} shows clear variation among the mid-IR multi-epoch bins, both in the individual magnitudes and in its color.  
The galaxy's changing color reveals that it can be classified as a mid-IR AGN (i.e. W1$-$W2 $\geqslant 0.5$) in 19\% of the epochs explored ($p_{AGN,0.5}=19$).
Thus, looking for ``color variability'' reveals ``previously hidden'' AGN activity that is detected only by multi-epoch observations.
As discussed further in Section \ref{subsec:pAGN}, such ``hidden'' AGN activity is found in more than 1,200 galaxies.

In Figure~\ref{fig:GIF}, we illustrate this phenomenon in another way for another void galaxy, SDSS J102912.86+245720.4, where we show the multi-epoch W1$-$W2 colors as measured in the averaged bins of the NEOWISE and MEP data. This object's classification as a mid-IR AGN with W1$-$W2 $\geqslant 0.8$ or W1$-$W2 $\geqslant 0.5$ appears for $p_{\text{AGN, 0.8}} \approx 20$\%, or $p_{\text{AGN, 0.5}} \approx 90$\% of its observed time respectively, or for $p_{\text{AGN, Mateos}} \approx 40$\%  displaying colors within the \citet{mateos_using_2012} wedge; the single epoch AllWISE measurement, which is shown here as the blue star, would have dismissed it as an AGN based on the W1$-$W2 $\geqslant 0.8$ or the Mateos et al. criteria.  

We present a detailed discussion of the statistics of $p_{\text{AGN, 0.5}}$ and $p_{\text{AGN, 0.8}}$ among the samples of void and wall galaxies in Section~\ref{subsec:pAGN}.

\section{Defining Variable Sources} 
\label{sec:vardefs}

We have presented so far the definition and individual potential of three main metrics for identifying variable sources: the Pearson coefficient \pr that statistically quantifies the coupling between variations in the W1 and W2 bands, \sig that measures the combined significance of variances in W1 and W2, and $p_{\rm AGN}$ as the fraction of observed time with mid-IR AGN colors.
Because ultimately we are interested in comparing the degree to which void and wall galaxies harbor AGN activity, as measured by variability in their mid-IR emission, we explore here ways in which we can define the most complete and efficient variability criteria based on individual or combinations of the metrics we have calculated so far.

\begin{figure}
    \centering
    \includegraphics[width=1.0\linewidth]{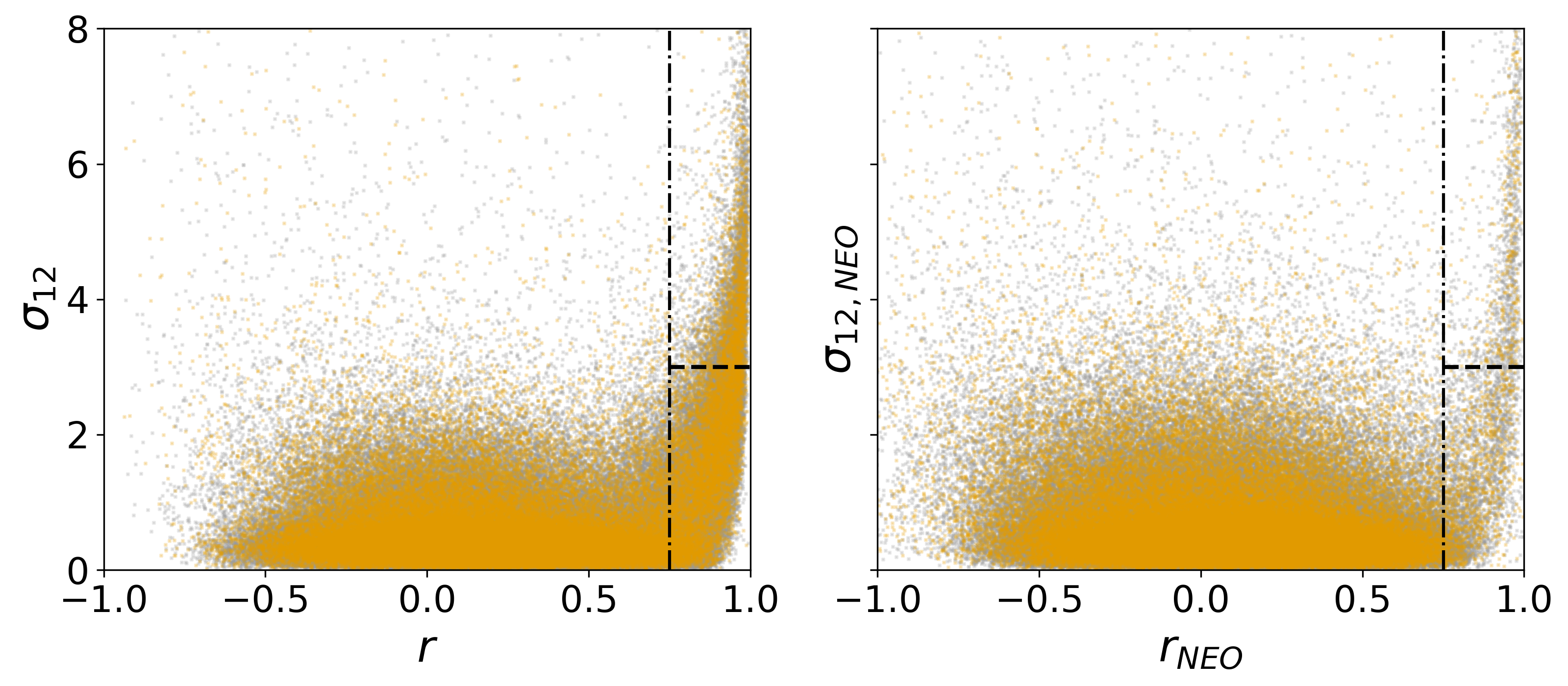}
    \caption{\sig versus Pearson \pr values for all void galaxies (orange dots) and wall galaxies (gray dots) when both metrics are calculated using all epochs of mid-IR data (left panel) and when they are calculated using only the NEOWISE epochs (right panel). Our conservative variability criterion is $r > 0.75$ (vertical black dash-dot line) and $\sigma_{12} > 3$ (black horizontal dashed line) for both calculation methods (both panels).}
    \label{fig:Sig12_vs_Pr}
\end{figure}

We explore four variability classes that incorporate all of the variability metrics discussed above,
along with single-epoch mid-IR colors for identifying AGNs:
\begin{enumerate}

\item A simple definition of variability based on the Pearson \pr coefficient: $r > 0.75$ and $r_{\rm NEO} > 0.75$ (hereafter $r_{\rm both} > 0.75$);

\item A more conservative definition of variability, where we combine the \pr and \sig variability metrics: $r > 0.75$, $r_{\rm NEO} > 0.75$, $\sigma_{12} > 3$  and $\sigma_{12, {\rm NEO}} > 3$ (hereafter $r_{\rm both} > 0.75$ and $\sigma_{12, {\rm both}} > 3$);

\item A color-variability definition based on the fraction of observed time that galaxies show W1 -- W2 colors consistent with mid-IR AGN behavior (i.e., W1 -- W2 $\geqslant 0.5$, and more stringently $\geqslant 0.8$), $p_{\rm AGN, 0.5} > 50$ or $p_{\rm AGN, 0.8} > 50$, depending on the color cut;

\item A final, comprehensive mid-IR AGN definition selecting all galaxies that satisfy at least one of the stringent AGN definitions we discuss in this paper:
    \begin{enumerate}
    
    \item $r_{\rm both} > 0.75$ and $\sigma_{12, {\rm both}} > 3$, or
     
    \item $p_{\rm AGN, 0.8} > 50$, or
    
    \item $W1-W2 \geqslant 0.8$.
    
    \end{enumerate}
\end{enumerate}

These criteria are rigorous because we enforce cuts in metrics calculated using both all light-curve epochs and only the NEOWISE epochs to avoid flagging galaxies as variable based solely on the brightness change during the hiatus between their MEP and NEOWISE measurements.
In particular, our second variability definition (combining $r_{\rm both} > 0.75$ and $\sigma_{12, {\rm both}} > 3$) closely follows the variability criteria of past mid-IR AGN variability studies \citep{polimera_morphologies_2018, kozlowski_quasar_2016, kozlowski_mid-infrared_2010}. Figure~\ref{fig:Sig12_vs_Pr} illustrates the complementarity of our variability criteria for \pr, $r_{\rm NEO}$, \sig, and $\sigma_{12, {\rm NEO}}$, confirming that sources with strong coupling between brightness changes in their W1 and W2 light curves (high \pr) tend also to display significant brightness variations (high \sig), thus making them promising variable candidates.

In the following subsections, we investigate a novel way to quantify the link between mid-IR variability and AGN behavior, i.e., the expectation that AGN-dominated galaxies become redder (larger W1$-$W2 colors) when brighter (lower W1 values), as well as the complementarity of our variability definitions, including potential brightness biases.

\subsection{Validating variability metrics: Quantifying the Redder-when-Brighter Behavior}

One way of quantifying the potential redder-when-brighter behavior of the objects selected as variable sources by our combinations of \pr, \sig, and $p_{\rm AGN}$ variability metrics is by designing a new Pearson correlation coefficient, $r_{\text{color, W1}}$, to measure the coupling of changes in W1 and W1$-$W2 over each galaxy's light curve (as opposed to simply between the W1 and W2 values), according to the same equations (\ref{eqn:r})--(\ref{eqn:vm}). With this definition, perfectly anti-correlated W1 and W1$-$W2 light curves result in a correlation coefficient of $r_{\text{color, W1}} = -1$ meaning that the galaxy exhibits redder colors (larger W1$-$W2 values) at brighter (smaller) magnitudes, or a {\it redder-when-brighter} trend that is expected with an increasing dominance of an AGN component over its host galaxy light.
In turn, a coefficient of $r_{\text{color, W1}} = 1$ means that the W1 and W1$-$W2 light curves are perfectly correlated (i.e., bluer-when-brighter trends), and $r_{\text{color, W1}} = 0$ reflects no correlation between W1 and the color. Thus, we should expect to see more prevalent redder-when-brighter behavior (indicated by $r_{\text{color, W1}} \sim -1$) among the most promising mid-IR variable candidates.

\begin{figure}
    \centering
    \includegraphics[width=1.0\linewidth]{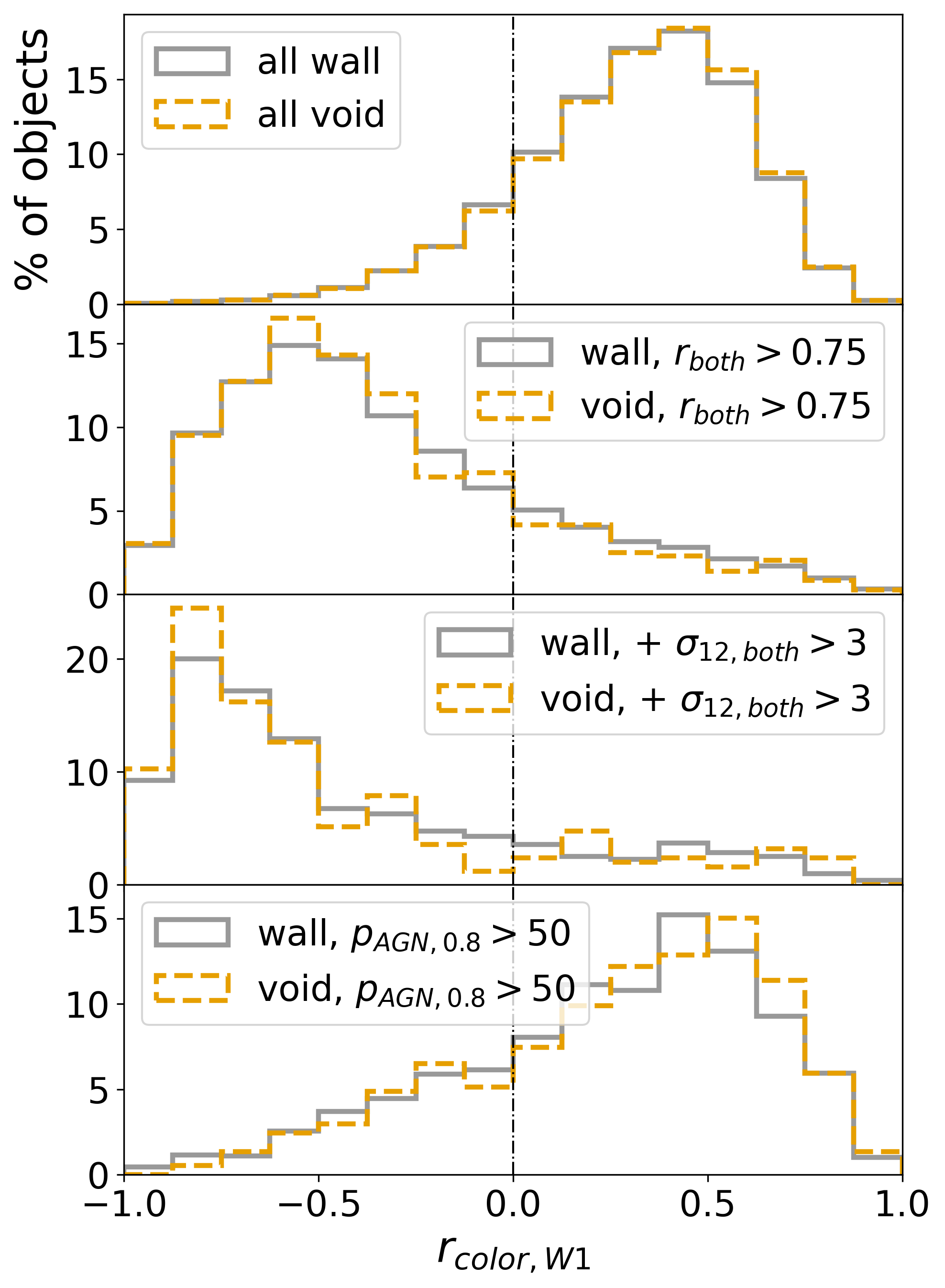}
    \caption{Distributions of the Pearson \pr correlation coefficient calculated between the epochs of W1 and W1$-$W2 for wall (gray solid line) and void (orange dashed lines) galaxies. The distributions for the whole void and wall galaxy samples are shown in the top panel, while the lower panels show the distribution for galaxies meeting different variability criteria: $r_{\rm both} > 0.75$,
    the combination of $r_{\rm both} > 0.75$ and $\sigma_{12, {\rm both}} > 3$, and $p_{{\rm AGN}, 0.8} > 50$. The black vertical dash-dot line divides galaxies that show bluer-when-brighter colors (i.e., $r_{\text{W1, color}} > 0$, right of the line) from the ones that show the redder-when-brighter colors (i.e., $r_{\text{W1, color}} < 0$, left of the line) that are consistent with increasing AGN dominance over the host galaxy light emission.}
    \label{fig:Pr_color}
\end{figure}

Figure~\ref{fig:Pr_color} presents these new 
$r_{\text{color, W1}}$ calculations, for the entire samples of void and wall galaxies as well as for subsamples of galaxies that are defined as variable based on various combination criteria for \pr, \sig, and $p_{\rm AGN}$.  

As expected, the vast majority of objects in the void and wall samples (top panel) have $r_{\text{color, W1}} > 0$, or non-AGN behavior, which is consistent with the likelihood that most galaxies do not host AGN; among the entire void and wall samples, only $\sim$14\% and 15\% show redder-when-brighter behavior, i.e., $r_{\text{color, W1}} < 0$.  However, as we enforce variability cuts based on $r_{\rm both}$ (second panel from the top) and combine these cuts with additional requirements for $\sigma_{12, {\rm both}}$ (third panel), the distribution becomes strongly skewed towards negative values of $r_{\text{color, W1}}$.
Among the galaxies that meet our strictest variability metric criteria ($r_{\rm both} > 0.75$ and $\sigma_{12, \text{both}} > 3$), the redder-when-brighter fraction is dramatically higher: about 81\%
of both void and wall galaxies have $r_{\text{color, W1}} < 0$. 
These trends support the validity and complementarity of our variability definition based on both \pr and \sig variability metrics because it indirectly selects for
galaxies that exhibit the redder-when-brighter trends expected with AGN activity.

Interestingly, the galaxies defined as variable based on their high fraction of time in which they exhibit mid-IR AGN colors ($p_{\rm AGN} > 50$) reveal a distribution
that remains skewed towards values of $r_{\text{color, W1}} > 0$. As we discuss in the following section, this subset of variable galaxies tends to also be skewed toward quite faint mid-IR emission, thus perhaps reducing the significance of the redder-when-brighter effect.

\subsection{Comparing \pr and \sig Metrics with $p_{AGN}$: A Brightness Bias}
\label{subsec:bias}

\begin{figure}
    \centering
    \includegraphics[width=1.0\linewidth]{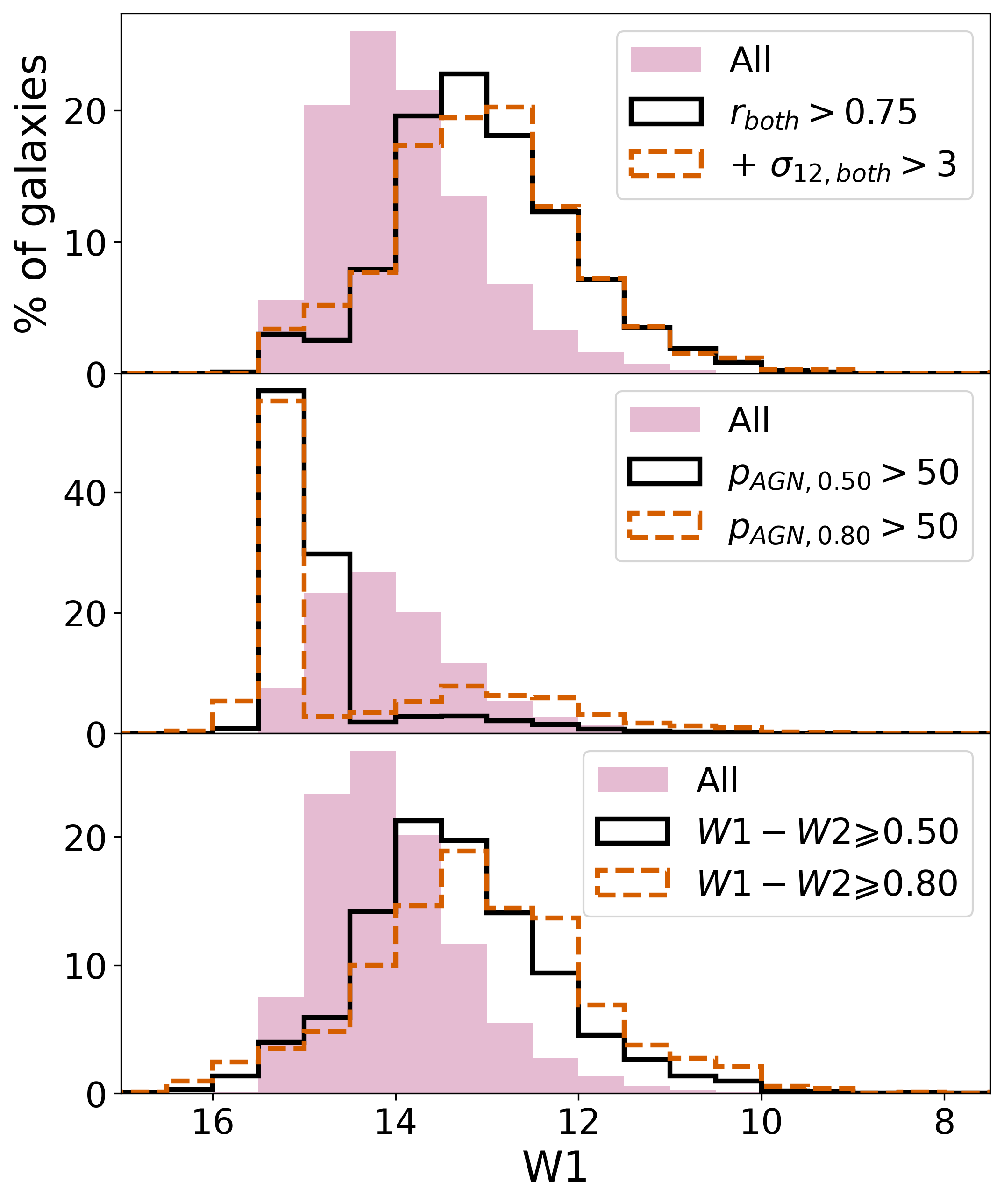}
    \caption{Distributions of W1 for the combined void and wall galaxy samples that meet different AGN candidate criteria based on \pr and \sig variability metrics (top), color variability (middle), and traditional single-epoch color cuts (bottom).}
    \label{fig:lum_bias}
\end{figure}

In order to investigate potential brightness biases introduced by the mid-IR variability selections we consider here, we look into 
the W1 dependence of the distributions of fractions of galaxies that meet various variability criteria as well as traditional AGN definitions.  Figure~\ref{fig:lum_bias} illustrates a comparison of the W1 distributions for samples defined by the metric variability based on $r_{\rm both} > 0.75$ and $\sigma_{12, {\rm both}} > 3$ (upper panel), the color variability based on $p_{{\rm AGN}, 0.5} > 50$ and $p_{{\rm AGN}, 0.8} > 50$ (middle panel) and single-epoch mid-IR color cuts (bottom panel).

Interestingly, we find that different variability criteria lead to different distributions in brightness.
The  W1 distributions of the fractions of galaxies that obey just $r_{\rm both} > 0.75$, as well as the combination of $r_{\rm both} > 0.75$ and $\sigma_{12, {\rm both}} > 3$, are statistically consistent with one another; a Kolmogorov-Smirnov (KS) test cannot reject the null hypothesis ($p \approx 0.87$) that the underlying probability distributions for the W1 values are identical.
Both distributions are shifted towards brighter hosts relative to the whole parent population of galaxies with no constraint on their multi-epoch emission behavior. The W1 distribution of galaxies meeting traditional mid-IR single-epoch color cuts (bottom panel) is somewhat similar, also consistent with objects that are brighter than the overall parent sample.

\begin{figure}
    \centering
    \includegraphics[width=1.0\linewidth]{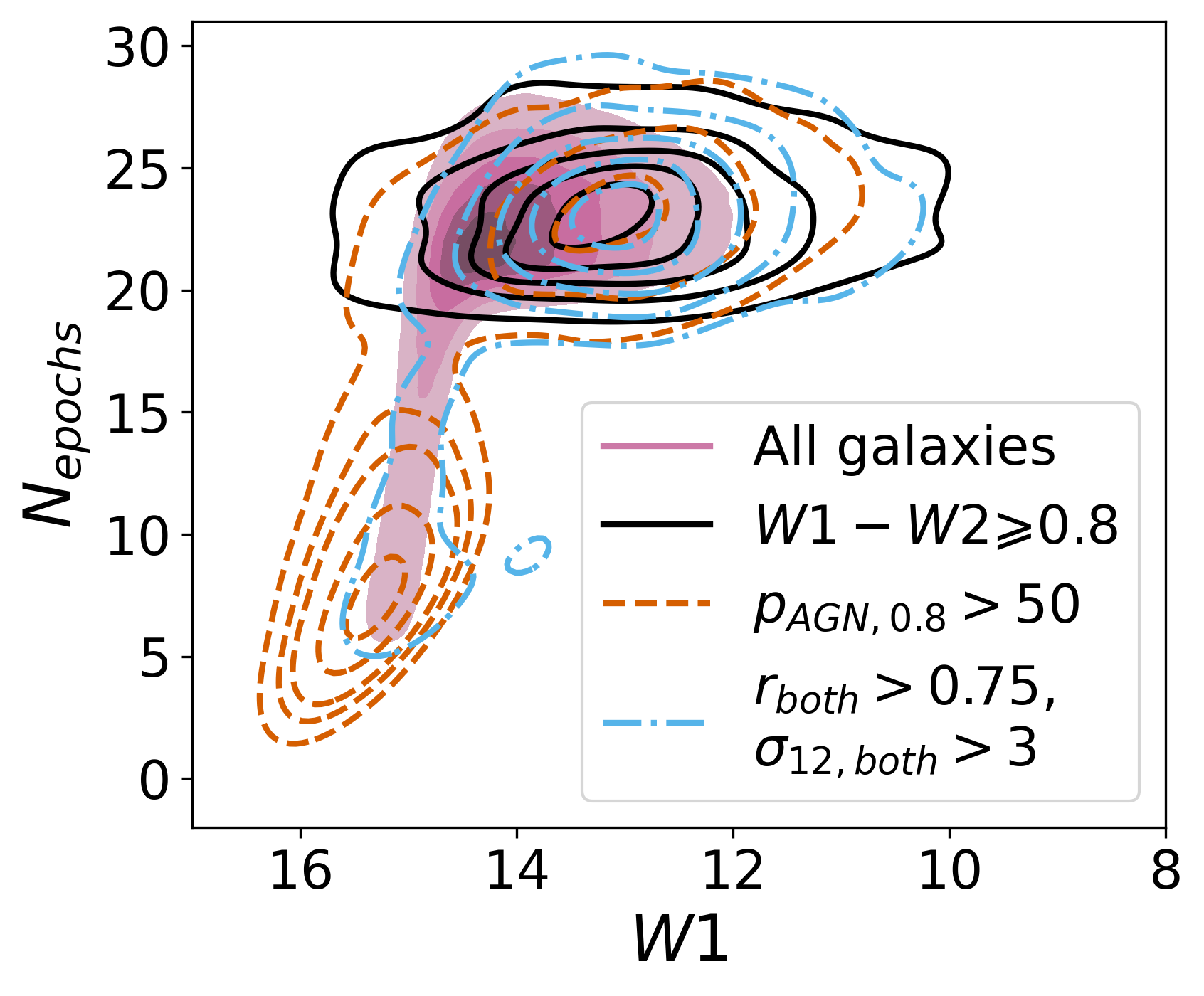}
    \caption{Number of light curve epochs (``bins") versus W1 apparent magnitude for the combined void and wall galaxy samples (pink filled contours) as well as the subsets that meet various cut for each mid-IR AGN diagnostic: red mid-IR colors W1$-$W2 $\geqslant 0.8$ (red dashed contours), majority of observed time with such AGN-like mid-IR colors ($p_{{\rm AGN}, 0.8} > 50$, orange dash-dot contours), and high variability metrics (i.e., $r_{\rm both} > 0.75$ and $\sigma_{12, {\rm both}} > 3$). The density contours correspond to factors of $n$ of the total number of objects in each class, where $n =$ 0.1, 0.2, 0.3, 0.5, 0.7, and 0.9 starting from innermost contour.}
    \label{fig:N_bias}
\end{figure}

However, the W1 distributions of galaxies meeting our color variability ($p_{\rm AGN} > 50$) criteria show a noticeably different trend, with a clear bias towards the faintest galaxies.
A plausible explanation for this brightness bias is the possibility that faint galaxies also tend to have fewer light curve epochs and therefore can satisfy the color variability criteria ($p_{\rm AGN} > 50$) by chance with fewer multi-epoch data.
Figure \ref{fig:N_bias} shows the distribution of the number of light curve epochs ($N_{\rm epochs}$) as a function of W1 for the combined void and wall galaxy sample (pink filled contours) as well as for the subsets of this combined sample that satisfy various cuts for each mid-IR AGN diagnostic considered: red single-epoch mid-IR colors (W1$-$W2 $\geqslant 0.8$; black continuous contours), majority of observed time with such AGN-like mid-IR colors ($p_{{\rm AGN},0.8} > 50$; red dashed contours), and high variability metrics ($r_{\rm both} > 0.75$ and $\sigma_{12, {\rm both}} > 3$; dot-dashed blue contours). It is quite apparent here that the high $p_{\rm AGN}$ galaxies show a strong bias not only towards fainter galaxies but also towards galaxies with fewer light-curve epochs, compared to the full galaxy sample.

Figure \ref{fig:N_bias} also reveals that there are biases associated with the other two AGN diagnostic methods. The traditional single-epoch red color criterion W1$-$W2 $\geqslant 0.8$ tends to select AGN candidates in galaxies with large amounts of multi-epoch data; the vast majority of AGN candidates selected using this method have more than $\sim$20 light curve epochs.  In contrast, AGN candidates selected using variability metrics \pr and \sig  match well the distributions of $N_{\rm epoch}$ and W1 associated with that of the full galaxy sample.

In summary, it is apparent that different criteria for variability uncover different AGN populations, characterized by rather distinct, and quite complementary host brightnesses.
While mid-IR variability defined via the \pr and \sig metrics uncover brighter samples of AGNs that are generally consistent with those defined by mid-IR single-epoch color criteria, employing multi-epoch color variability with $p_{\rm AGN}$ reveals a trove of faint AGN candidates, likely previously elusive to other AGN classifications.

\section{Results} \label{results}
\label{sec:results}

\subsection{AGN Fraction by \pr and \sig Variability Metrics} \label{subsec:AGNmetrics}

\begin{deluxetable*}{lcccc} 
\tablewidth{0 pt} 
\tablecaption{Variability Metric Results}
\tablehead{
\colhead{Sample} & \multicolumn{2}{c}{$r_{\rm both} > 0.75$} & \multicolumn{2}{c}{$r_{\rm both} > 0.75, \sigma_{\rm 12, both} > 3$}\\
\colhead{} & \multicolumn{2}{c}{\% (count)} & \multicolumn{2}{c}{\% (count)} \\
\cline{2-3} \cline{4-5}
\colhead{} & \colhead{out of Sample} & \colhead{out of Blue} & \colhead{out of Sample} & \colhead{out of Blue}
}
\startdata 
 Void & 1.60 ± 0.05 (1084) & 1.07 ± 0.04 (715) & 0.37 ± 0.02 (253) & 0.20 ± 0.02 (133) \\ 
 Wall & 1.63 ± 0.03 (3401) & 1.08 ± 0.02 (2238) & 0.40 ± 0.01 (844) & 0.180 ± 0.009 (372) \\ 
\\ 
Bright void & \textbf{2.31 ± 0.09 (749)} & 1.48 ± 0.07 (473) & 0.52 ± 0.04 (170) & \textbf{0.27 ± 0.03 (85)} \\ 
Bright wall & \textbf{2.14 ± 0.04 (2650)} & 1.41 ± 0.03 (1721) & 0.50 ± 0.02 (620) & \textbf{0.20 ± 0.01 (240)} \\ 
\\ 
Main void & 0.94 ± 0.05 (324) & 0.68 ± 0.04 (233) & 0.23 ± 0.03 (80) & 0.13 ± 0.02 (46) \\ 
Main wall & 0.87 ± 0.03 (740) & 0.61 ± 0.03 (511) & 0.26 ± 0.02 (220) & 0.15 ± 0.01 (130) \\ 
\\ 
Dwarf void & 1.6 ± 0.5 (11) & 1.4 ± 0.5 (9) & 0.4 ± 0.3 (3) & 0.3 ± 0.2 (2) \\ 
Dwarf wall & 2.2 ± 0.7 (11) & 1.3 ± 0.5 (6) & 0.8 ± 0.4 (4) & 0.4 ± 0.3 (2)
\enddata
\label{tbl:mega}
\tablecomments{Fractions of void and wall galaxy samples and their luminosity subsamples that satisfy the respective variability metrics. The listed errors are $\pm 1 \sigma$ Poisson uncertainties. The few pairs of fractions that do not overlap within their uncertainties (highlighted in bold) indicate a higher fraction of variable sources within voids than within walls. For each sample, we present the variable fraction out of that sample and just out of the blue galaxies (W1$-$W2 $< 0.5$) within that sample. } 
\end{deluxetable*}

We present and compare here the fractions of mid-IR variable galaxies in our void and wall samples measured via the Pearson's \pr (Section~\ref{pearsonr}) and \sig (Section~\ref{sigma12}) metrics, both calculated using all light curve epochs and just the NEOWISE epochs.  The results are summarized in 
Table~\ref{tbl:mega}.

We find that when using a simple definition of variability based only on Pearson's \pr values (with both \pr and $r_{\rm NEO} > 0.75$), the only pair of void and wall AGN fractions that do not overlap within $\pm 1 \sigma$ Poisson uncertainties belongs to the Bright sample, with voids appearing to harbor a slightly higher fraction of variable galaxies than walls (significant at the $1.9 \sigma$ level). This difference remains apparent even for more stringent criteria for variability, i.e., $r_{\rm both} > 0.85$ or $r_{\rm both} > 0.95$. Once we require the stricter variability definition based on both statistical metrics  $r_{\rm both} > 0.75$ and $\sigma_{12,{\rm both}} > 3$, we see no statistically significant differences between the void and wall fractions of variable sources within the whole samples or other luminosity sub-samples.

We also compare here the fraction of mid-IR variable sources within just the ``Blue" (i.e. single-epoch W1$-$W2 $< 0.5$) void and wall galaxies, for all these variability measures. Among these subsamples, once again the void Bright galaxies host a higher fraction of variable sources than their wall counterparts, when the stricter variability definition (combination of \pr and \sig) is used. This difference is significant at the $2.4 \sigma$ level and suggests that the variability metrics \pr and \sig are slightly more efficient at ``uncovering'' hidden AGN in void environments.
 
To summarize, when defined based on \pr and \sig metrics, the variability fractions for the overall samples of void and wall galaxies remain statistically indistinguishable and therefore do not provide a clear distinction of the level of AGN activity between the two environments.  Nevertheless, the differences exhibited by the Bright luminosity sub-sample point towards a possibly stronger presence of AGNs within the void regions.  Further comparisons of these fractions with those derived from other AGN/variability definitions are presented in the next subsections.

\subsection{AGN Fraction by Color Variability} \label{subsec:pAGN}

\begin{deluxetable*}{lcccc} 
\tablewidth{0 pt} 
\tablecaption{\% AGN Within the Multi-epoch Data}
\tablehead{
\colhead{Sample} & \multicolumn{2}{c}{$p_{{\rm AGN},0.5} > 50$} & \multicolumn{2}{c}{$p_{{\rm AGN},0.8} > 50$} \\
\colhead{} & \multicolumn{2}{c}{\% (count)} & \multicolumn{2}{c}{\% (count)} \\
\cline{2-3} \cline{4-5}
\colhead{} & \colhead{out of Sample} & \colhead{out of Blue} & \colhead{out of Sample} & \colhead{out of Blue}
}
\startdata 
 Void & 10.3 ± 0.1 (6957) & 9.2 ± 0.1 (6138) & 1.09 ± 0.04 (738) & 0.66 ± 0.03 (440) \\ 
 Wall & 7.20 ± 0.06 (15064) & 6.23 ± 0.06 (12877) & 0.75 ± 0.02 (1564) & 0.40 ± 0.01 (834) \\ 
\\ 
Bright void & 4.3 ± 0.1 (1391) & 2.9 ± 0.1 (920) & 0.53 ± 0.04 (171) & 0.12 ± 0.02 (37) \\ 
Bright wall & 2.91 ± 0.05 (3596) & 1.73 ± 0.04 (2105) & 0.42 ± 0.02 (525) & 0.071 ± 0.008 (87) \\ 
\\ 
Main void & 15.4 ± 0.2 (5333) & 14.6 ± 0.2 (5014) & 1.51 ± 0.07 (524) & 1.12 ± 0.06 (384) \\ 
Main wall & 13.2 ± 0.1 (11276) & 12.6 ± 0.1 (10609) & 1.17 ± 0.04 (992) & 0.85 ± 0.03 (721) \\ 
\\ 
Dwarf void & 35 ± 3 (233) & 32 ± 3 (204) & 6 ± 1 (43) & 3.0 ± 0.7 (19) \\ 
Dwarf wall & 38 ± 3 (192) & 34 ± 3 (163) & 9 ± 1 (47) & 5 ± 1 (26)
\enddata
\label{tbl:pAGN}
\tablecomments{Fractions of void and wall galaxy samples and their luminosity subsamples who spend more than 50\% of their time (defined as number of epochs in their light curves) displaying mid-IR AGN-like colors, i.e., W1$-$W2 $\geqslant 0.5$ ($p_{{\rm AGN},0.5} > 50$) and $\geqslant 0.8$ ($p_{{\rm AGN},0.8} > 50$). The listed errors are $\pm 1 \sigma$ Poisson uncertainties. With the exception of Dwarf galaxies, 
all of the void samples show significantly ($\geqslant 2.3 \sigma$) higher AGN fractions.} 
\end{deluxetable*} 

When we compare the fractions of void and wall galaxies that spent more than half of their WISE/NEOWISE observed time exhibiting  AGN-like mid-IR colors, i.e., via $p_{{\rm AGN},0.5}$ and $p_{{\rm AGN},0.8}$ parameters, 
we find a clear trend that void galaxies in the whole, Bright, and Main samples are more likely than wall galaxies to host AGN-like activity.
When we define AGN via mid-IR colors W1$-$W2 $\geqslant 0.5$ ($p_{{\rm AGN},0.5}$), this result is significant at the $24 \sigma$, $12 \sigma$, and $9.5 \sigma$ levels for the full, Bright, and Main galaxy samples, respectively; when AGNs are defined by W1$-$W2 $\geqslant 0.8$ mid-IR colors ($p_{{\rm AGN},0.8}$), this result is significant at the $8.5 \sigma$, $2.5 \sigma$, and $5.2 \sigma$ levels for the full, Bright, and Main galaxy samples, respectively.
The Dwarf sample may simply be too small for these tests. We present in Table~\ref{tbl:pAGN}, the fractions of void and wall galaxies with $p_{{\rm AGN},0.5}$ and $p_{{\rm AGN},0.8} > 50$ out of the whole samples as well as for their Bright, Main and Dwarf luminosity sub-samples, along with the corresponding fractions calculated out of the Blue galaxies showing single-epoch mid-IR colors of W1$-$W2 $< 0.5$, which therefore have not been previously identified as AGNs.

When examining the fractions of void vs. wall galaxies in the ``out of Blue'' subsets in Table~\ref{tbl:pAGN}, it is apparent that the high $p_{\rm AGN}$ objects are significantly more prevalent among blue void galaxies than among blue wall galaxies as well; these trends hold again for the whole sample ($24 \sigma$ significance for $p_{{\rm AGN},0.5}$; $8.1 \sigma$ significance for $p_{{\rm AGN},0.8}$), Bright sample ($12 \sigma$ significance for $p_{{\rm AGN},0.5}$; $2.3 \sigma$ significance for $p_{{\rm AGN},0.8}$), and Main sample ($9.3 \sigma$ significance for $p_{{\rm AGN},0.5}$; $4.6 \sigma$ significance for $p_{{\rm AGN},0.8}$). This indicates that the color variability method is more efficient at ``uncovering'' AGNs in void environments than in wall environments. This is consistent with our finding that color variability recovers AGN activity in fainter sources, given that void galaxies tend to be smaller, fainter, and bluer \citep{rojas_photometric_2004, hoyle_luminosity_2005, ceccarelli_large-scale_2008, ceccarelli_impact_2022}. Since these properties also generally make void AGN harder to identify, our results suggest that color variability is an important tool for identifying void AGN.
When we define AGN-like colors with the standard mid-IR cut-off of W1$-$W2 $\geqslant 0.5$, we find 19,000 objects showing multi-epoch color variability despite their blue single-epoch colors.
When using the more conservative color threshold of W1$-$W2 $\geqslant 0.8$, the number of AGN uncovered is more than 1,200.

We also note that the tendency of higher $p_{\rm AGN}$ among the void galaxies holds regardless of how conservative the color cut for AGN-definition is, or whether we compare fractions for time thresholds that are smaller than 50\%; the differences remain strong in favor of more prevalent AGN activity in voids versus walls within the whole, Bright and Main subsamples even when we use thresholds of  $p_{\rm AGN} >$ 25\%, 20\%, or 10\%.

\subsection{AGN Fraction by Variability Metrics and Color}

\begin{figure*}
    \centering
    \includegraphics[width=1.0\linewidth]{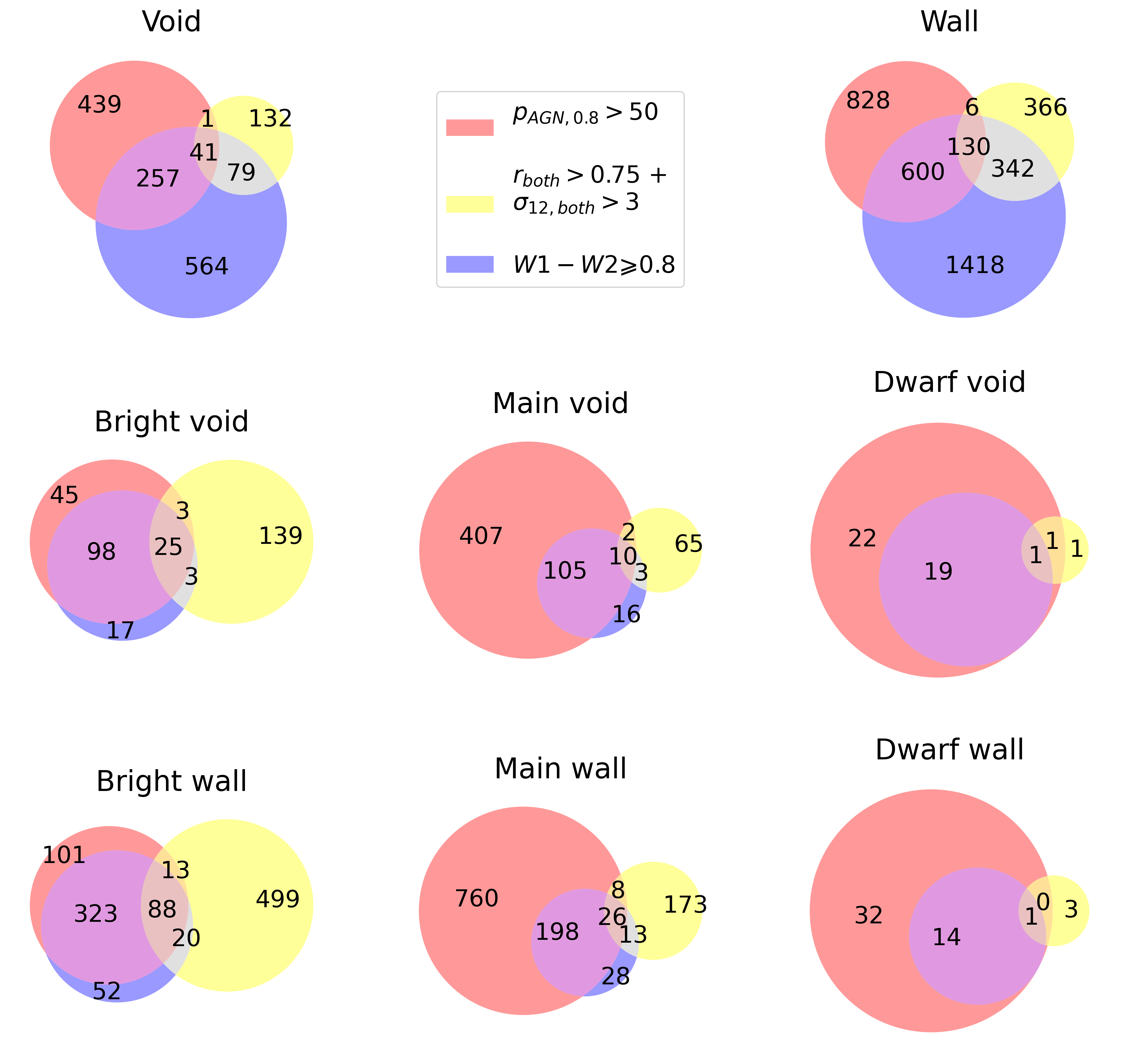}    
    \caption{Venn diagrams for the counts of void and wall mid-IR AGN candidates that fulfill the most stringent combinations of cuts for three AGN diagnostics: variability metrics ($r_{\rm both} > 0.75$ and $\sigma_{12, {\rm both}} > 3$; yellow), percent of time displaying AGN-like mid-IR colors ($p_{{\rm AGN}, 0.8} > 80$; red), and single-epoch mid-IR colors (W1$-$W2 $\geqslant 0.8$; blue). Top row shows results for the whole samples of void (left) and wall (right) galaxies, and the next two rows indicate numbers for the Bright, Main and Dwarf luminosity subsamples within the voids (middle row) and walls (bottom row). The circle areas are scaled relative to the total number of galaxies shown in each diagram, i.e., the number of galaxies that satisfy any one of the three AGN criteria.}
    \label{fig:venn}
\end{figure*}

Using calculations of the variable fractions for void and wall galaxies using the strictest thresholds for each diagnostic method, we now compare the relative sizes of the AGN-variable candidate populations with AGN fractions derived from single-epoch mid-IR colors. Figure~\ref{fig:venn} illustrates this comparison via Venn diagrams for the whole samples of void and wall sources, as well as for the Bright, Main and Dwarf luminosity subsamples; the listed numbers indicate the individual sample sizes, while the circle dimensions scale with the relative size of each sample out of the total number of galaxies that obey any of the three AGN and variability definitions considered here.

Overall, the relative numbers, fractions, and cross fractions of AGN candidates identified by the three methods appear relatively similar in voids and walls. Additionally, across both the void and wall samples it is clear that galaxies selected for high $r_{\rm both}$ and $\sigma_{12, {\rm both}}$ (yellow circles) show relatively little overlap with the galaxies that spend the majority of observed time displaying AGN-like mid-IR colors (i.e., $p_{\rm AGN} > 50$; red circles), or with galaxies that are traditionally considered mid-IR AGN due to their single-epoch W1$-$W2 red colors (blue circles). Thus, the variability metrics and the color variability method each identify unique populations of AGN candidates, and both methods show potential to identify hundreds of mid-IR AGN that were previously overlooked based only on their single-epoch colors.

Like with the overall samples of void and wall galaxies, the fractions and cross fractions of AGN candidates by the three different AGN selection methods are relatively similar between void and wall hosts of all brightnesses. 
Nevertheless, several interesting trends become apparent: within the Bright samples (top row), requiring high $r_{\rm both}$ and $\sigma_{12, {\rm both}}$ (yellow circles) identifies the largest fraction of AGN candidates, in contrast to the $p_{\rm AGN}$ method (red circles) which identifies the smallest fraction of candidates. These trends reverse for the lowest luminosity sub-samples; the $p_{\rm AGN}$ criterion selects the largest fraction of candidates among the Dwarf wall and void samples, while the \pr and \sig variability metric method contributes the smallest fraction. This comparison reveals once again that the sensitivity of these mid-IR AGN variability selection cuts
varies by host galaxy luminosity. Of the three selection methods displayed here, the variability metric \pr and $\sigma_{12}$ cuts seem to be the most sensitive for Bright host galaxies, while the color variability $p_{\rm AGN}$ cut is most sensitive for identifying the least luminous mid-IR AGN candidates, very much consistent with the trends discussed before (Section \ref{subsec:bias}), as exhibited in Figure ~\ref{fig:lum_bias}. 

Interestingly, the cross-fractions between selection methods also scale with luminosity. In particular, we see changes by luminosity in the relative fractions of galaxies not identified as AGN candidates by their single-epoch mid-IR colors (outside of the blue circles) but which are uncovered as AGNs by variability metrics (yellow circles) or color variability (red circles): i) among the most luminous void and wall galaxies, the largest fraction of previously hidden mid-IR AGN candidates are exposed by variability metric cuts (yellow); ii) in contrast, the vast majority of Bright AGN candidates identified by color variability (red) are already identified using traditional single-epoch color cuts (blue), resulting in a large cross-fraction (purple). While this purple cross-fraction remains significant across the entire luminosity range, the color variability method, rather than variability metrics, uncovers the largest fraction of previously overlooked AGN candidates among the less luminous Main and Dwarf sources. 

These trends once again indicate the complementarity of these two methods: variability metrics \pr and \sig are most sensitive to exposing previously overlooked AGN in the most luminous host galaxies, while color variability tends to uncover AGN candidates in less luminous hosts.

\subsection{AGN Fraction by the Comprehensive mid-IR AGN Definition}
\label{subsec:comp}

To accomplish our ultimate goal of comparing the mid-IR AGN fraction in void and wall environments, we combine the unique AGN candidate populations identified by the color variability, variability metrics, and traditional color AGN diagnostic methods via the comprehensive variability criterion.  To recap, this definition selects the AGN candidates that meet the most stringent version of at least one of the three aforementioned diagnostic methods: i) $r_{\rm both} > 0.75$ and $\sigma_{12, {\rm both}} > 3$, or ii) $p_{\rm AGN, 0.8} > 50$, or iii) W1$-$W2 $\geqslant 0.8$ (see Section \ref{sec:vardefs} for details).
We find that $\sim$1\% of our whole galaxy sample meets this comprehensive definition, which is consistent with AGN fractions found by past studies of mid-IR AGN variability \citep{kozlowski_mid-infrared_2010, kozlowski_quasar_2016, polimera_morphologies_2018}.

\begin{figure}
    \centering
    \includegraphics[width=1.0\linewidth]{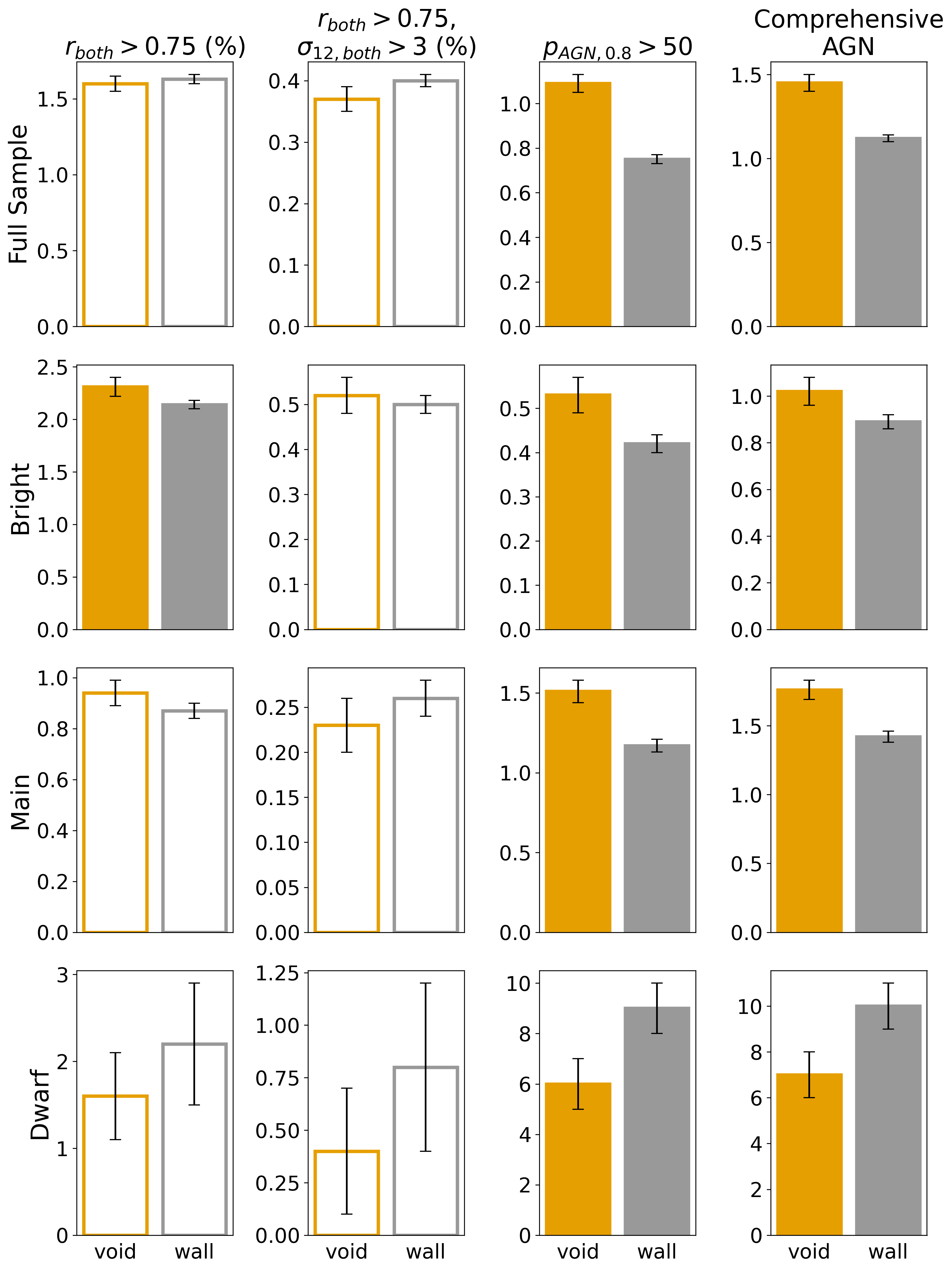}
    
    \caption{Fractions of void (orange) and wall (gray) galaxies that satisfy the following AGN diagnostics: the simple definition of variability, $r_{\rm both} > 0.75$, ({\it leftmost column}), the addition of $\sigma_{12, {\rm both}} > 3$ to the simple definition, ({\it second column from the left}), majority of observed time displaying AGN-like mid-IR colors, $p_{{\rm AGN}, 0.8} > 50$, ({\it second column from the right}), and the comprehensive mid-IR AGN  candidate definition (Section \ref{sec:vardefs}, {\it rightmost column}).
    The bars are filled with color when the void and wall fractions do not overlap within their shown $\pm 1 \sigma$ Poisson error bars, and are unfilled otherwise.}
    \label{fig:bartable}
\end{figure}

For a different, more encompassing view of these results, we present a summary of the comparison of the fractions of AGN candidates found through all these methods individually, together with the comprehensive method,  in the bar chart of Figure \ref{fig:bartable}.
The fractions among wall galaxies are shown by the bars on the right of each pair, while bars for the void fractions are on the left, with the filled bars indicating the void and wall fractions that do not overlap within their $\pm 1 \sigma$ Poisson error bars.
The conclusion is that the mid-IR AGN candidate fraction is generally higher in void galaxies than in their wall counterparts within the whole galaxy samples, as well as the Bright and Main subsamples. This result, utilizing the comprehensive AGN definition, is significant at the $7.0 \sigma$, $2.3 \sigma$, and $4.7 \sigma$ levels for the full, Bright, and Main galaxy samples, respectively. Although not shown, we point out that the void AGN candidate fractions are also higher among the Blue subsets, at a significance level $7.7 \sigma$ for the full blue void and wall samples, and $\geqslant 3.3 \sigma$ for the Bright and Main luminosity subsamples. While the trends seem reversed among the Dwarf sources, with AGN activity appearing more prevalent in the denser large scale environments, we caution that the number statistics are very small for this luminosity range.

To conclude, we employ here for the first time mid-IR variability to find evidence for a larger AGN fraction among galaxies in the most underdense large-scale environments compared to their counterparts in wall regions.
Given the comparison of the individual contributions of each method to the comprehensive definition of variability, as discussed above, it is rather clear that this overall trend towards more frequent AGN activity within void environments is not skewed by a certain way of defining variability.
Our finding of a higher AGN fraction in voids is consistent with past studies of the AGN fraction in voids that employ different AGN definitions and discovery methods, like optical emission line diagnostics \citep[e.g.,][]{constantin_active_2008} and single-epoch mid-IR color diagnostics \citep[e.g.,][]{ceccarelli_impact_2022}.


\section{Conclusions and Discussion} \label{conclusions}
\label{sec:conclusion}

In this work, we have quantified and compared for the first time the mid-IR variability, and therefore the AGN activity of galaxies of vastly different environmental properties, i.e., residing in cosmic voids and walls, that are matched in their optical brightness.  

To recap, we define AGN candidate sources via four definitions for mid-IR variability: (i and ii) two metric variability methods that employ the Pearson $r > 0.75$ coefficient alone as well as in combination with $\sigma_{12} > 3$, where \pr quantifies the coupling between brightness variations in the W1 and W2 mid-IR wavelength bands, and \sig quantifies the significance of these variations compared to galaxies of similar apparent magnitudes; (iii) the color variability method that employs $p_{\rm AGN} > 50$, where $p_{\rm AGN}$ is the percentage of a galaxy’s light curve epochs that show red mid-IR colors, i.e., W1$-$W2 $\geqslant 0.8$; and (iv) a most comprehensive mid-IR AGN definition selecting all galaxies that satisfy at least one of the most stringent variability and color definitions listed above (see Section \ref{sec:method} for detailed definitions of all of these parameters).

The following are our main conclusions:
\begin{enumerate}

\item We find clear evidence for a larger mid-IR variable-AGN fraction among high and moderate-luminosity void galaxies compared to their wall counterparts. When using a comprehensive mid-IR AGN definition factoring in both color and metric variability, this result is significant at the $7.0 \sigma$ level for the full void and wall galaxy samples. Color variability ($p_{\rm AGN}$) alone reveals a strong trend in this direction (significant at the $24 \sigma$ level for the full void and wall samples using $p_{\rm AGN, 0.5}$), while metric variability (\pr and \sig) alone identifies statistically indistinguishable AGN candidate fractions between void and wall hosts, regardless of host luminosity.

\item Identifying true variability using the combination of the AllWISE Multiepoch Photometry (MEP) dataset and NEOWISE multi-epoch photometry datasets requires variability metrics calculated using the NEOWISE dataset alone (e.g., $r_{12, {\rm NEO}}$ and $\sigma_{12, {\rm NEO}}$) because otherwise metrics may flag light curves that only reflect significant variability during the ~3-year gap between the MEP and NEOWISE measurements. While these gaps are acknowledged by past studies \citep[e.g.,][]{guidry_WD_2024, son_mid-infrared_2022, secrest_low_2020}, ours is the first study of AGN to utilize separate consideration of NEOWISE data to validate variability candidates.

\item We introduce for the first time a statistical tool for quantifying the link between variability and the redder-when-brighter behavior, which reveals a luminosity bias as well as complementarity in identifying variable AGN candidates via the aforementioned variability criteria:
the metric variability method shows the most sensitivity for identifying AGN candidates among the most luminous hosts while the color variability method is most sensitive among low-luminosity galaxy hosts. 

\item Each of the metric-based and the color-based measures of mid-IR variability identifies unique populations of previously missed AGNs; both methods are showing potential to reveal hundreds of mid-IR AGN candidates among blue hosts as defined based on their single-epoch colors (i.e., W1$-$W2 $< 0.5$). The fractions of the newly identified mid-IR variable-AGNs among blue host galaxies are higher among the voids (e.g. at the $7.7 \sigma$ significance level when using the comprehensive mid-IR AGN definition for the full blue void and wall samples), suggesting once again a more prolific AGN activity in the most underdense large scale structures of the universe.

\end{enumerate}

Why would void galaxies prefer stronger mid-IR variability and thus AGN activity? Past studies have presented several possible explanations for mechanisms that could result in this trend.
Based on optical line emission properties alone, \citet{constantin_active_2008} suggest that their findings of subtle differences between wall and void AGN activity may be explained by a longer/slower duty cycle for the void galaxies in their evolutionary sequence from star-forming to AGN-dominance to weakly accretion and star-formation nuclear activity.
Alternatively, using a Principal Component Analysis method of identifying the independent effects of distinct aspects of source environment on the triggering of nuclear activity \citep{sabater_interact_2013}, \citet{miraghaei_void_2020} suggest that a higher void AGN fraction could be due to an elevated level of one-on-one galaxy interactions in these most underdense environments, where galaxy mergers are more ubiquitous than in denser environments due to lower velocity dispersions \citep{sureshkumar_mergers_2024}.
Additionally, \citet{ceccarelli_impact_2022} suggest that gas flowing from void interiors outwards could preferentially fuel the central black holes of void hosts, while such a fuel source would be deficient within denser regions thanks to interactions with cluster or group hot gas.

While these scenarios are not mutually exclusive, our results seem to favor the explanation that a higher AGN fraction in voids is driven by elevated levels of one-on-one galaxy interactions.  More frequent one-on-one interactions would drive frequent episodes of star formation, which is consistent with past studies that find higher specific star formation rates and a higher fraction of star-forming galaxies among voids \citep{Rojas05, hoyle_luminosity_2005, ceccarelli_large-scale_2008, constantin_active_2008}, while also driving morphological distortions and propelling gas towards galaxy centers to fuel AGN activity \citep{constantin_active_2008}. Thus, these processes would preferentially favor the formation of obscured AGN, with elevated levels of morphological distortion that would match the population of mid-IR AGN identified via either single-epoch red colors \citep[e.g.,][]{ellison_definitive_2019} or variability \citep[][]{polimera_morphologies_2018}, like we find here.
Further simulations of galaxy interactions in voids and  morphological studies of NEOWISE-selected mid-IR variable galaxies are needed to corroborate this explanation.

Our finding of an elevated AGN fraction within voids at first seems to conflict with past studies that observe similar AGN fractions in void regions compared to wall regions \citep{liu_spectral_2015} or rich galaxy clusters \citep{amiri_role_2019}. However, both of these previous studies rely solely on optical emission line AGN diagnostics, which miss strongly obscured AGN \citep[e.g.][]{amiri_role_2019}. It has been shown in numerous studies that obscured AGN elusive to optical emission line techniques can be recovered using mid-IR diagnostics, including mid-IR variability \citep[e.g.,][]{polimera_morphologies_2018, secrest_low_2020, ceccarelli_impact_2022}, and we reveal here that the fractions of such elusive AGN are clearly higher among voids than in walls, especially among blue mid-IR galaxies.

Lastly, it is important to note that mid-IR variability metrics like the ones we employ in this study have the potential to detect non-AGN types of sources, such as supernovae (SNe) and tidal disruption events (TDEs). For example, \citet{myers_sne_2024} use NEOWISE data to discover one transient in an emerging family of core-collapse SNe whose mid-IR emission outshines their optical emission, and \citet{masterson_tdes_2024} uncovered 12 candidates of dust-obscured TDEs missed by optical and X-ray surveys by seeking sources with fast-rising NEOWISE light curves. While these discoveries demonstrate the broader utility of mid-IR variability, they also show that detections of non-AGN variables is unlikely to hinder our study’s conclusions, as their numbers remain well within the realm of the errors associated with the numbers involved in our comparative analysis. \citet{polimera_morphologies_2018} estimate an upper limit of just 6\% for the fraction of supernovae among the mid-IR variable galaxies they selected using $\sigma_{12} \geqslant 3$ and  $r \geqslant 0.8$, and \citet{dodd_notdes_2023} find that the large majority of mid-IR outbursts in galaxies are consistent with AGN activity rather than obscured TDEs.

Ultimately, our study is the first to employ mid-IR variability to probe nuclear activity in galaxies inhabiting the most underdense environments, the cosmic voids. Our findings not only confirm the putative important effects of large-scale structure on galaxy evolution but also demonstrate the promise of mid-IR variability as a prime tool for future studies of AGN activity.

\begin{acknowledgements}

Funding for the Sloan Digital Sky Survey (SDSS) and SDSS-II has been provided by the Alfred P. Sloan Foundation, the Participating Institutions, the National Science Foundation, the U.S. Department of Energy, the National Aeronautics and Space Administration, the Japanese Monbukagakusho, and the Max Planck Society, and the Higher Education Funding Council for England. The SDSS Web site is http://www.sdss.org/.

The SDSS is managed by the Astrophysical Research Consortium (ARC) for the Participating Institutions. The Participating Institutions are the American Museum of Natural History, Astrophysical Institute Potsdam, University of Basel, University of Cambridge, Case Western Reserve University, The University of Chicago, Drexel University, Fermilab, the Institute for Advanced Study, the Japan Participation Group, The Johns Hopkins University, the Joint Institute for Nuclear Astrophysics, the Kavli Institute for Particle Astrophysics and Cosmology, the Korean Scientist Group, the Chinese Academy of Sciences (LAMOST), Los Alamos National Laboratory, the Max-Planck-Institute for Astronomy (MPIA), the Max-Planck-Institute for Astrophysics (MPA), New Mexico State University, Ohio State University, University of Pittsburgh, University of Portsmouth, Princeton University, the United States Naval Observatory, and the University of Washington.

This work has made use of data products from the Wide-field Infrared Survey Explorer (WISE) and the SDSS. WISE is a joint project of the University of California, Los Angeles, and the Jet Propulsion Laboratory/California Institute of Technology, funded by the National Aeronautics and Space Administration.

This publication makes use of data products from the Wide-field Infrared Survey Explorer, which is a joint project of the University of California, Los Angeles, and the Jet Propulsion Laboratory/California Institute of Technology, and NEOWISE, which is a project of the Jet Propulsion Laboratory/California Institute of Technology. WISE and NEOWISE are funded by the National Aeronautics and Space Administration.

This research has made use of the NASA/IPAC Infrared Science Archive, which is funded by the National Aeronautics and Space Administration and operated by the California Institute of Technology.  
We are grateful for the valuable comments from an anonymous reviewer, which helped to improve this paper.
A. A. acknowledges the Robertson Scholars Leadership Program for their support. This work was partially supported by the National Science Foundation award NSF: AST \#1814594.

\facility {IRSA, NEOWISE, Sloan, WISE}

\end{acknowledgements}
\bibliographystyle{aasjournal}
\bibliography{Mid-IR-AGN}{}
\end{document}